\documentclass[10pt,journal,compsoc]{IEEEtran}
\ifCLASSOPTIONcompsoc
  \usepackage[nocompress]{cite}
\else
  \usepackage{cite}
\fi

\usepackage{graphicx}
\usepackage{balance}
\usepackage{comment}

\usepackage{booktabs}
\newcommand{\ra}[1]{\renewcommand{\arraystretch}{#1}}

\usepackage[ruled]{algorithm2e}
\usepackage{authblk}
\usepackage{verbatim} 
\usepackage{graphics} 
\usepackage{epsfig} 
\usepackage{epstopdf}
\usepackage{subfigure}
\usepackage{balance}
\usepackage{comment}
\usepackage{booktabs}
\usepackage{amsmath}
\usepackage{amssymb}
\usepackage{amsfonts}
\usepackage{bm}
\usepackage{colortbl}
\usepackage{tabularx}
\usepackage{color}
\usepackage{xspace}
\usepackage{graphicx}    
\usepackage{url}         

\usepackage{algpseudocode}
\usepackage{placeins}
\usepackage{multirow}
\usepackage{caption}
\usepackage{enumitem}
\usepackage{leading}
\usepackage{pgfplots}
\usepackage{tikz}
\usepackage{etex}
\usepackage{authblk}
\usepackage{caption}
\usepackage{collcell}
\usepackage{hhline}
\usepackage{pgf}

\newcommand\gray{gray}

\newcommand\ColCell[1]{%
	\pgfmathparse{#1<.8?1:0}%
	\ifnum\pgfmathresult=0\relax\color{white}\fi
	\pgfmathparse{1-#1}%
	\expandafter\cellcolor\expandafter[%
	\expandafter\gray\expandafter]\expandafter{\pgfmathresult}#1}

\newcolumntype{E}{>{\collectcell\ColCell}c<{\endcollectcell}}

\newcommand{\RN}[1]{%
	\textup{\uppercase\expandafter{\romannumeral#1}}%
}

\author{Dong Ma, \textit{Student Member,~IEEE,} Guohao Lan, \textit{Member,~IEEE,} Weitao Xu, \textit{Member,~IEEE,} \\ Mahbub Hassan, \textit{Senior Member,~IEEE,} and Wen Hu, \textit{Senior Member,~IEEE,}
\thanks{Dong Ma is with the School of Computer Science and Engineering, University of New South Wales (UNSW), Sydney, Australia, and also with the Data61, CSIRO, Eveleigh, Australia (E-mail: dong.ma1@unsw.edu.au).}
\thanks{Guohao Lan is with the Department of Electrical and Computer Engineering, Duke University, Durham, NC 27708, USA (E-mail: guohao.lan@duke.edu). The work was done when the author was with UNSW.}
\thanks{Weitao Xu is with the Department of Computer Science, City University of Hong Kong, Hong Kong (e-mail: weitaoxu@cityu.edu.hk). The work was done when the author was with UNSW.}
\thanks{Mahbub Hassan is with the School of Computer Science and Engineering, University of New South Wales (UNSW), Sydney, Australia (E-mail: mahbub.hassan@unsw.edu.au).}
\thanks{Wen Hu is with the School of Computer Science and Engineering, University of New South Wales (UNSW), Sydney, Australia (E-mail: wen.hu@unsw.edu.au).}

}

\begin{document}

\title{Simultaneous Energy Harvesting and Gait Recognition using Piezoelectric Energy Harvester\\
}

\maketitle

\begin{abstract}
Piezoelectric energy harvester, which generates electricity from stress or vibrations, is gaining increasing attention as a viable solution to extend battery life in wearables. Recent research further reveals that, besides generating energy, PEH can also serve as a passive sensor to detect human gait power-efficiently because its stress or vibration patterns are significantly influenced by the gait. However, as PEHs are not designed for precise measurement of motion, achievable gait recognition accuracy remains low with conventional classification algorithms. The accuracy deteriorates further when the generated electricity is stored simultaneously. To classify gait reliably while simultaneously storing generated energy, we make two distinct contributions. First, we propose a preprocessing algorithm to filter out the effect of energy storage on PEH electricity signal. Second, we propose a long short-term memory (LSTM) network based classifier to accurately capture temporal information in gait-induced electricity generation. We prototype the proposed gait recognition architecture in the form factor of an insole and evaluate its gait recognition as well as energy harvesting performance with 20 subjects. Our results show that the proposed architecture detects human gait with 12\% higher recall and harvests up to 127\% more energy while consuming 38\% less power compared to the state-of-the-art.

\end{abstract}

\begin{IEEEkeywords}
Piezoelectric Energy Harvesting, Simultaneous Energy Harvesting and Sensing, Gait Recognition, Deep Learning, LSTM
\end{IEEEkeywords}

\section{Introduction}
Piezoelectric energy harvester (PEH), which can harvest electrical energy from mechanical stress or vibration, has become an attractive solution to power many industrial sensor nodes~\cite{sudevalayam2011energy,vullers2010energy}, as well as wearable devices~\cite{bionic,SOLEPOWER,mide}. Interestingly, because PEH energy harvesting is influenced by the surrounding contexts, the same PEH can also serve as an energy-free sensor to detect a wide range of machine and human contexts, such as measuring the airflow of air conditioning systems~\cite{xiang2013powering}, detecting human activities~\cite{hassan2018kinetic}, demodulating acoustic communications~\cite{lan2017veh}, and many more as surveyed in~\cite{8944276}. 

Given that gait recognition is considered as an attractive option for human identification and authentication~\cite{gafurov2007survey}, our focus in this paper is to explore reliable gait recognition in wearable devices using PEH as a passive sensor. A recent attempt by Xu et al.~\cite{xuTMCGait} revealed that although PEH can detect gait 82\% more power-efficiently than conventional sensors, i.e., accelerometers, it cannot match the high recognition accuracy achievable with accelerometers. A further limitation of the work in~\cite{xuTMCGait} is that the authors used the PEH only as a sensor, but the generated electricity was not harvested in a capacitor. Thus the interference caused by energy harvesting on the PEH-generated AC voltage and its resulting impact on gait recognition performance was not captured by the experiments in~\cite{xuTMCGait}. As a matter of fact, the impact of energy harvesting on the sensing performance of PEH was found to be significant, e.g., in airflow monitoring study~\cite{xiang2013powering}, which forced existing solutions to employ separate PEHs, i.e., one dedicated to sensing only, while the other is used for energy harvesting. As dedicated use of PEH increases hardware cost and complexity, we seek solutions that can realize simultaneous energy harvesting as well as reliable gait recognition using the same piece of PEH. Finally, the work in~\cite{xuTMCGait} considered gait recognition from PEH embedded in hand-held devices, which generated extremely small amount of power as walking-induced PEH vibrations have low energy density. Our goal is to design a PEH-based gait recognition system that can detect gaits reliably while generating sufficient energy that can be of practical use. 

The major contributions of this work are summarized as follows:
\begin{itemize}
	
	\item We design a novel simultaneous energy harvesting and gait recognition architecture, called simultaneous energy harvesting and sensing (SEHS). 
	In this architecture, we propose a classifier based on long short-term memory (LSTM) deep neural network to accurately capture the temporal information latent in gait-induced electricity generation. We further devise a preprocessing algorithm to filter out the effect of energy storage on the PEH electricity signal as much as possible before the signal is fed to the classifier.  
	
	\item We implement the proposed SEHS architecture using off-the-shelf PEH inside a shoe sole, which allows a large amount of energy generation from direct foot strikes during walking.  
	
	\item We evaluate the performance of the implemented SEHS prototype with 20 subjects. We observe that the proposed LSTM can learn the gait patterns 
	directly from the raw signals without filtering or preprocessing, whereas, existing classifiers struggle without filtering. With the proposed filtering, our results show that SEHS can detect human gait with 12\% higher recall and harvests up to 127\% more energy while consuming 38\% less power compared to the state-of-the-art presented in~\cite{xuTMCGait}. 
\end{itemize}

The rest of the paper is organized as follows. Section~\ref{section:SEHS} illustrates the effect of PEH energy harvesting on its sensing signal and presents the proposed filtering algorithm to minimize this effect. Design of LSTM neural networks for gait classification is presented in Section~\ref{s:gait_recognition}. Section~\ref{sec:evaluation} presents SEHS prototyping and its evaluation in terms of both gait recognition and energy harvesting performance. Current limitations and potential solutions are discussed in Section~\ref{sec:discuss}, followed by a review of related works in Section~\ref{section:related_work}. We conclude the paper in Section~\ref{section:conclusion}.

\section{SEHS Design}
\label{section:SEHS}

In this section, we present the design of the proposed SEHS architecture, illustrate the effect of energy harvesting on the sensing signal, and explain the proposed filtering algorithm to minimize the effect of energy harvesting on information sensing.

\begin{figure}[]
	\centering
	\subfigure[]{
		\centering
		\includegraphics[scale=0.63]{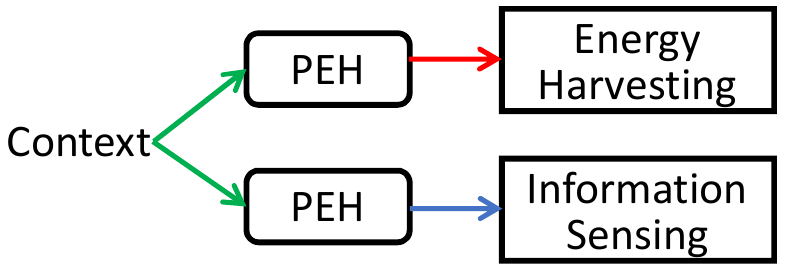}	
		\label{fig:arch1-separate}}
	\subfigure[]{
		\centering
		\includegraphics[scale=0.63]{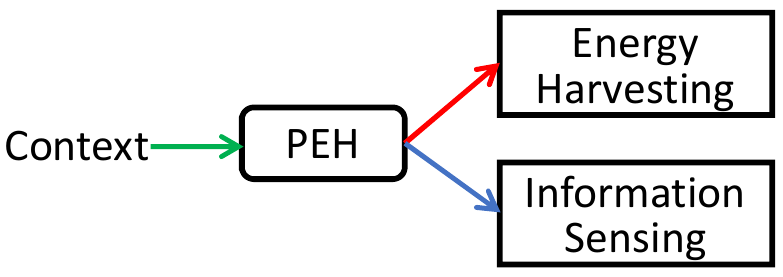}
		\label{fig:arch3-sehs}}
	\caption{(a) Requirement of separate PEHs for energy harvesting and sensing in existing architecture~\cite{xiang2013powering}, and (b) use of the same PEH for simultaneous energy harvesting and sensing in the proposed SEHS architecture.}
	\label{fig:architecture comparison}
	\vspace{-0.15in}
\end{figure}

\subsection{Limitations of Current Architecture}
Most prior works on PEH-based sensing \cite{kalantarian2015monitoring, xuTMCGait, lan2019entrans} only considered the sensing circuit without implementing energy harvesting and storage components, i.e., no capacitor was included in the circuit to store the harvested energy, where the main focus is to demonstrate various sensing capabilities of PEH. In~\cite{xiang2013powering}, Xiang et al. built a self-powered airflow monitoring system in which one PEH was utilized to sense the speed of airflow and the other to harvest energy from vibrations caused by the airflow. By using two PEHs, this approach completely avoids any interference from the interactions between harvesting and sensing. However, it increases system complexity, form factor, and cost.

\subsection{Design of the Proposed SEHS}
As shown in Figure~\ref{fig:architecture comparison}, the aims of the SEHS design are to (1) store the harvested energy, i.e., the rectified AC voltage generated by the PEH, in a capacitor, and (2) read the same AC voltage for context sensing using minimal power consumption. Unfortunately, reading the PEH AC voltage requires some additional processing, which would consume some power. Note that PEH generates electric potential proportional to the applied strain~\cite{poulin2004generation} and the polarization of the generated electricity corresponds to the direction of the induced deformation, producing the alternating voltage (AC). PEH usually generates an open-circuit AC voltage within minus decades volts to decades volts~\cite{bowen2014piezoelectric}, while the commonly used ADC (Analog-to-Digital Converter) can only measure the non-negative voltage ranging from 0V to 5V. Consequently, the voltage signal cannot be acquired when PEH output is directly connected to ADC. To address this problem, most existing works utilize an amplifier~\cite{xiang2013powering} or voltage divide circuit~\cite{xu2017gaitkey,lan2019entrans} to measure the AC voltage using a single ADC channel, which consumes several hundreds of $\mu W$. 

\begin{figure}[]
	\centering
	\includegraphics[scale = 0.76]{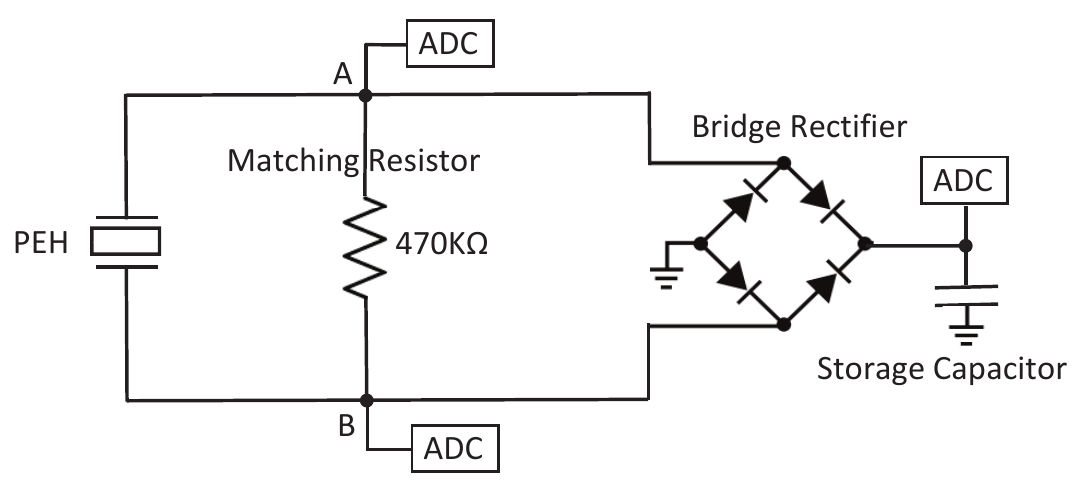}
	\caption{Circuit design of the proposed SEHS architecture.}
	\label{fig:circuit}
\end{figure}
\begin{figure*}[]
	\centering
	\begin{minipage}[]{0.66\textwidth}
		\centering
		\subfigure[]{
			\hspace{-0.3in}
			\includegraphics[scale=0.39]{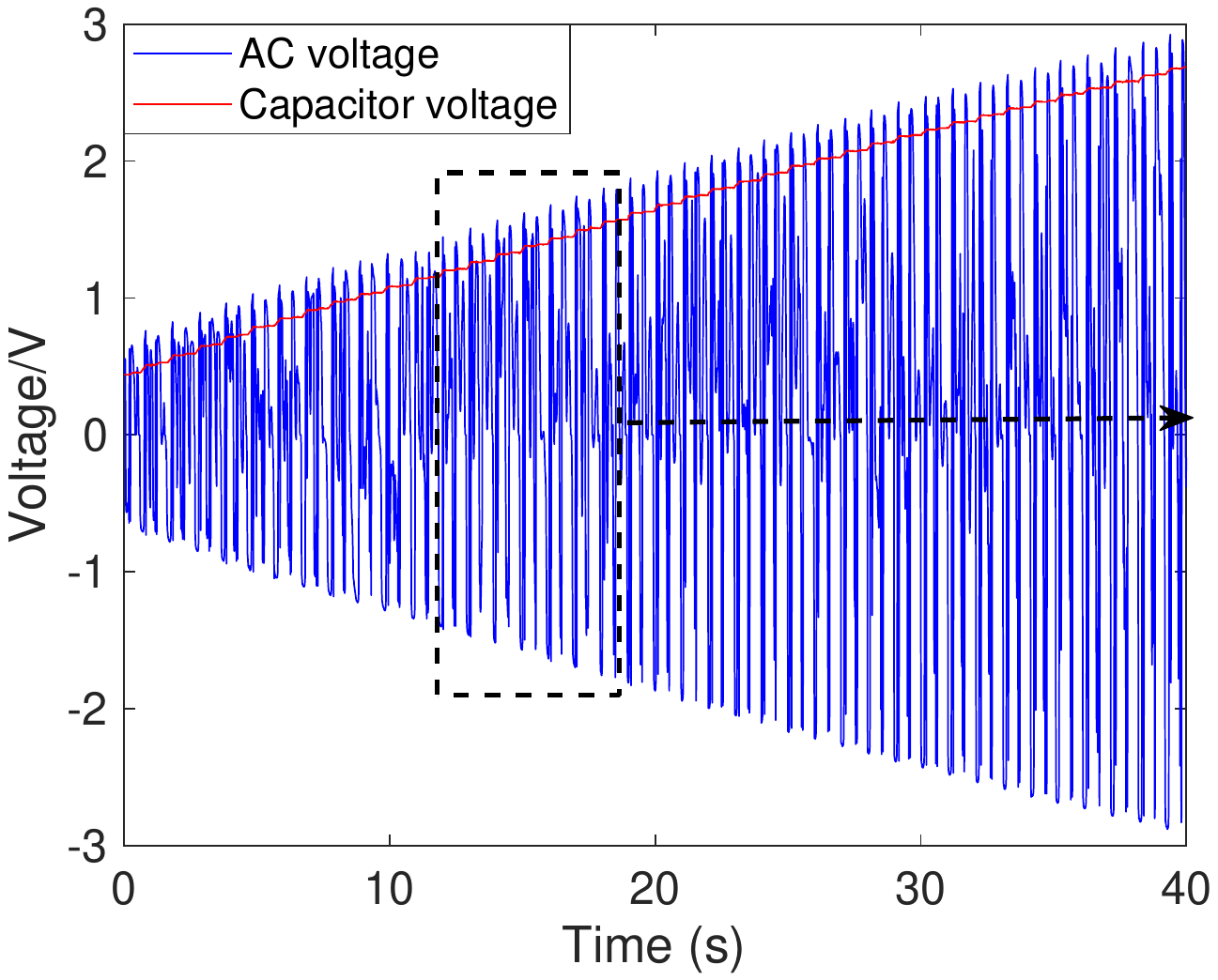}
		}
		\subfigure[]{
			\hspace{-0.2in}
			\includegraphics[scale=0.31]{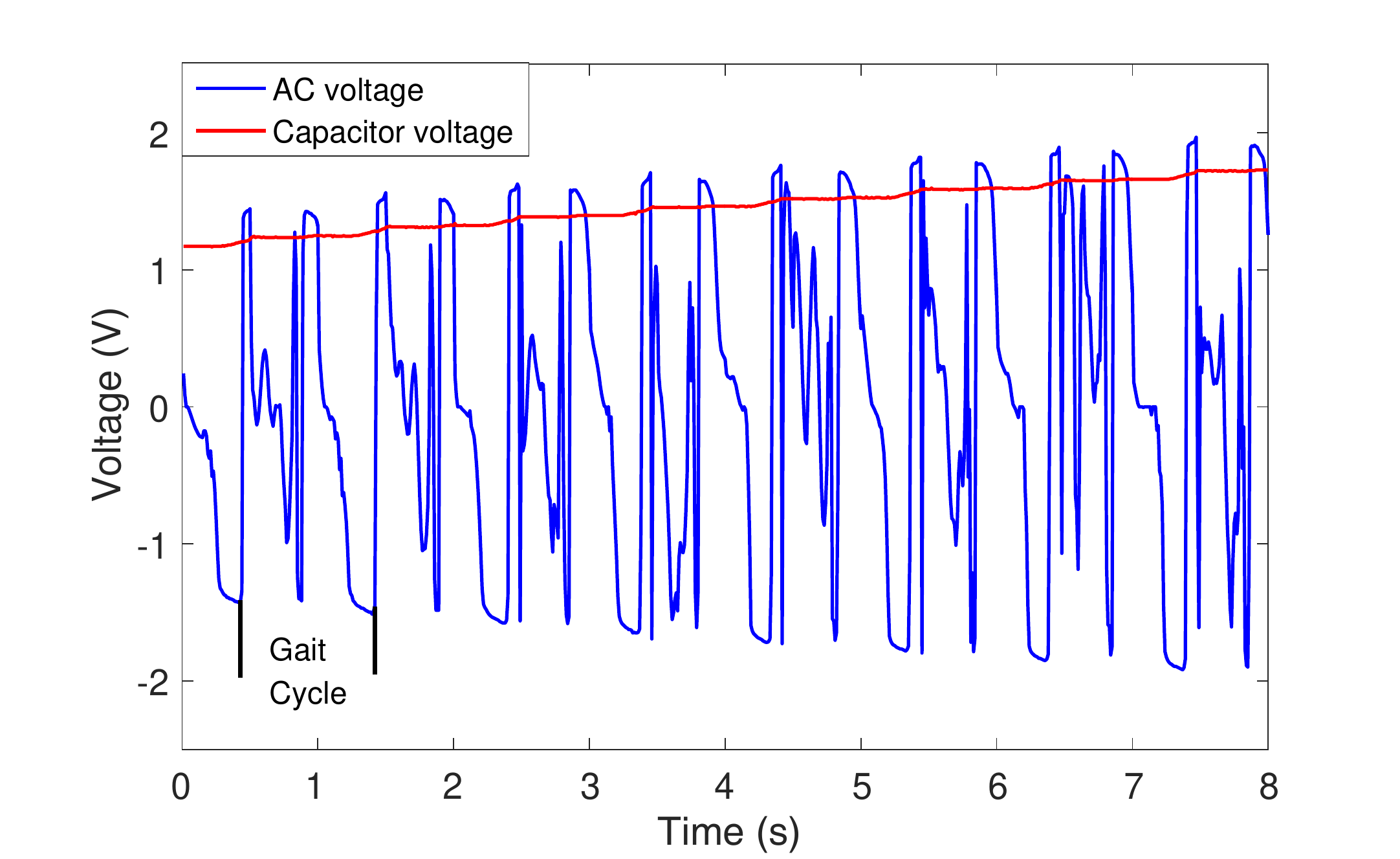}
		}
		\caption{(a) The original AC voltage and capacitor voltage and (b) its magnified version for a short period. }
		\label{fig:rawVolcompare}
	\end{minipage}
	\begin{minipage}[]{0.33\textwidth}
		\hspace{-0.4in}
		\includegraphics[scale=0.34]{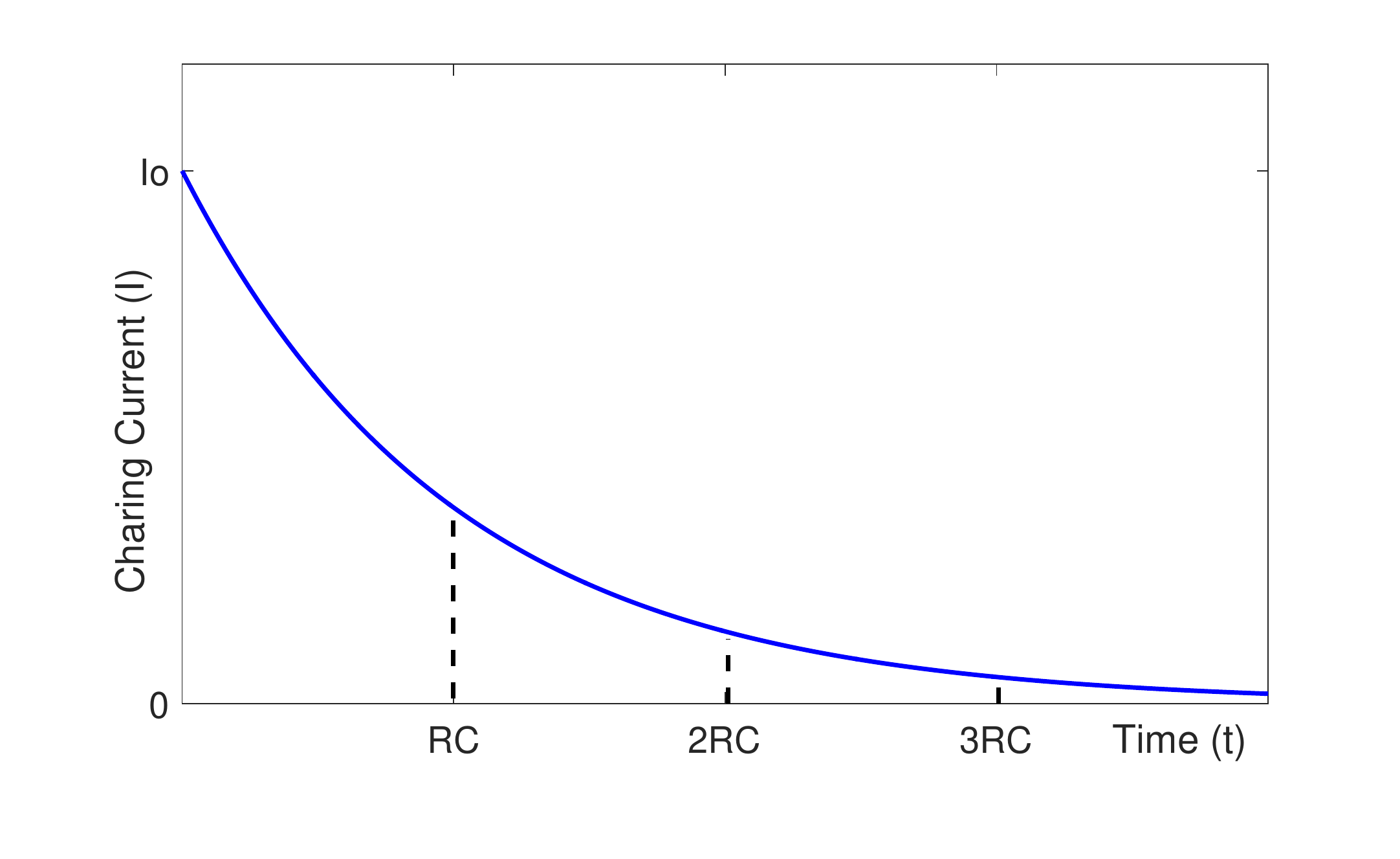}		
		\caption[Caption for LOF]{Capacitor charging current as a function of time. $RC$ is the time constant and $I_0$ is the initial current when capacitor voltage is 0.}
		\label{fig:theoCurrent}
	\end{minipage}
\end{figure*}

One of our design goals is to minimize the sensing related power consumption. We achieve this by trading off voltage divider with additional ADC channels, as ADC channels consume very little power on the order of 1-2 $\mu W$. Figure~\ref{fig:circuit} shows the circuit design of the proposed SEHS architecture. A matching resistor is used to limit the peak amplitude of the AC voltage within the ADC readable range. Instead of using a single ADC channel to capture the whole AC waveform, we use two ADC channels, i.e., point A ($V_A$) and point B ($V_B$), to measure the voltage on the matching resistor. Since the two points are directly connected to the output of PEH, the generated voltage, $V$, can be easily derived by subtracting the measured voltage at point B from point A (both are non-negative values), i.e., $V = V_A - V_B$. The energy flows to the capacitor through a full-bridge rectifier which is used to convert the AC voltage to DC voltage. With this circuit, the proposed SEHS architecture is able to collect the voltage signal and store the generated energy by using the same piece of PEH hardware.

\subsection{Effect of Energy Storage on Sensing}
\label{challenge}

Unfortunately, when a capacitor is used to store the harvested energy, its dynamic states (stored energy) modifies the current or AC voltage signal generated by the PEH. To illustrate the effect of energy harvesting on AC voltage reading, we collected 
data when our circuit was used in the shoe of a waking subject (see Section \ref{s:proto} for prototype details). Figure~\ref{fig:rawVolcompare} illustrates the waveform of the sampled AC voltage signal as well as the capacitor voltage. Intuitively, the peak-to-peak amplitude of the AC voltage should be approximately identical, as the entire trace is extracted from the same person while walking with a consistent style~\cite{xuTMCGait}. However, we can see that the peak-to-peak amplitude of the AC voltage is actually increasing with time as the capacitor voltage rises, i.e., more energy is being stored. The similar phenomenon is also described in~\cite{xiang2013powering}. 



The observed effect of energy harvesting on AC voltage in Figure~\ref{fig:rawVolcompare} can be explained as follows. Using capacitor theory~\cite{zhong2003current}, Figure~\ref{fig:theoCurrent} illustrates how the charging current decreases while the capacitor is being charged. This theory reveals that the current flow to the PEH is dropping when energy is being harvested and stored in the capacitor. Note that the PEH has a large internal resistance on the order of M$\Omega$~\cite{xiang2013powering, huang2016battery} and the output voltage is determined by the load resistance as well as the internal resistance~\cite{huang2016battery, sodano2004estimation}. With the current flow decreased, the voltage on the internal resistance of PEH is reduced. As a result, the amplitude of the output voltage is increased, which explains the increasing peak-to-peak amplitude of the AC voltage in Figure~\ref{fig:rawVolcompare}.


How to minimize the distortion of the sensing signal in a SEHS architecture remains an open problem. Note that the state-of-the-art in~\cite{xiang2013powering} did not actually solve this problem, but instead used two separate PEHs to avoid this issue. In the following subsection, we propose a filtering algorithm that can minimize the distortion effect of energy harvesting on the sensing signal. 

\subsection{Filtering Algorithm}
\label{section:Amplitude normalization}

\begin{algorithm}[b]
	\caption{Proposed Filtering Algorithm}
	\label{algorithm:filtering}
	\LinesNumbered
	\KwIn{$V_A(t), V_B(t), V_C(t), V^*$}
	\KwOut{$V_A^{'}(t)$, $V_B^{'}(t)$, $V^{'}(t)$}
	\bf Main procedure:\\
	\For{t = 1, 2, ..., N}{
		\eIf{$V_A(t) \geq V_C(t)$}{
			$V_A^{'}(t) = V_A(t) - V_C(t) + V^*$;
		}{
			$V_A^{'}(t) = V^* * (V_A(t) / V_C(t))$;
		}
		the same operation for $V_B(t)$;\\
		obtain $V_B^{'}(t)$;\\
		$V^{'}(t)$ = $V_A^{'}(t)$ - $V_B^{'}(t)$;	 
	}
\end{algorithm}

From Figure~\ref{fig:rawVolcompare}, we can observe that the main effect of storing energy is on the amplitude of AC voltage, where the amplitude continues to increase with increasing capacitor voltage. The aim of the filtering algorithm, therefore, is to prevent the increasing capacitor voltage from lifting the AC voltage without destroying the pattern of the signal. 
	Algorithm \ref{algorithm:filtering} shows the proposed filtering algorithm, where $V_A(t)$ and $V_B(t)$ represent the voltage at time $t$ measured at point A and B, respectively, as shown in Figure~\ref{fig:circuit} with $V_A^{'}(t)$ and $V_B^{'}(t)$ representing their corresponding filtered values. At time $t$, $V_C(t)$ is the capacitor voltage and $V^{'}(t)$ is the final AC voltage used for gait recognition. We introduce a constant, $V^*$, to compensate the impact of capacitor voltage. Particularly, when $V_A(t)$ (or $V_B(t)$) is higher than $V_C(t)$, the capacitor is being charged and its current voltage directly lifts the actual voltage at point A (or B). Thus, we obtain the difference between $V_A(t)$ and $V_C(t)$ and add it to $V^*$ (line 4). When $V_A(t)$ (or $V_B(t)$) is lower than $V_C(t)$, we multiply $V^*$ by the proportion of $V_A(t)$ and $V_C(t)$  to retain the gait patterns (line 6). Since the amplitude of the filtered signal is affected by the choice of $V^*$, we will evaluate the impact of $V^*$ on recognition performance in Section~\ref{s:filter performance}.
\begin{figure*}[]
	\centering
	
	\subfigure[UniLSTM]{
		\includegraphics[height=5.4cm]{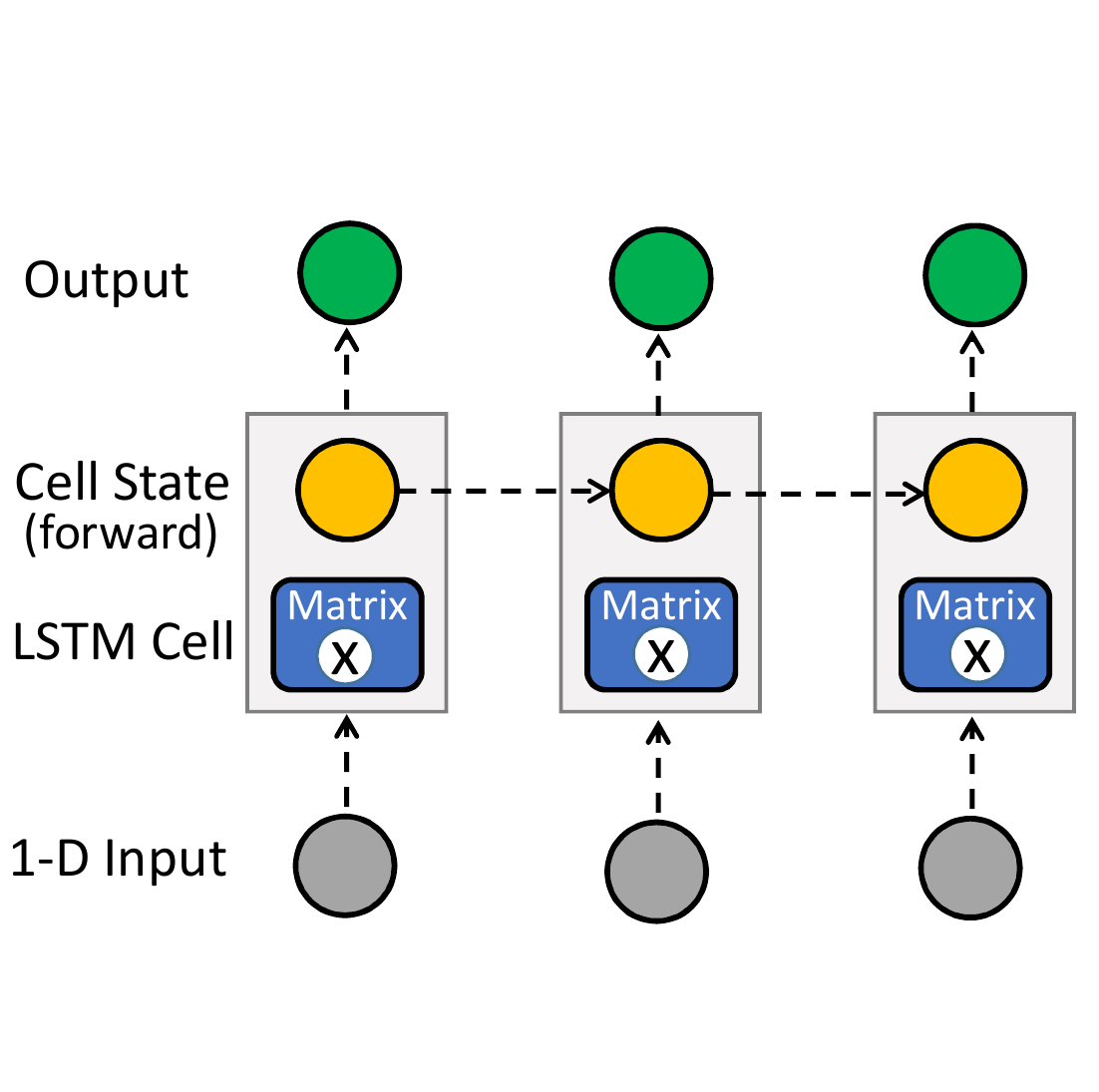}
		\label{fig:unilstm}}
	\subfigure[BiLSTM]{
		\includegraphics[height=5.4cm]{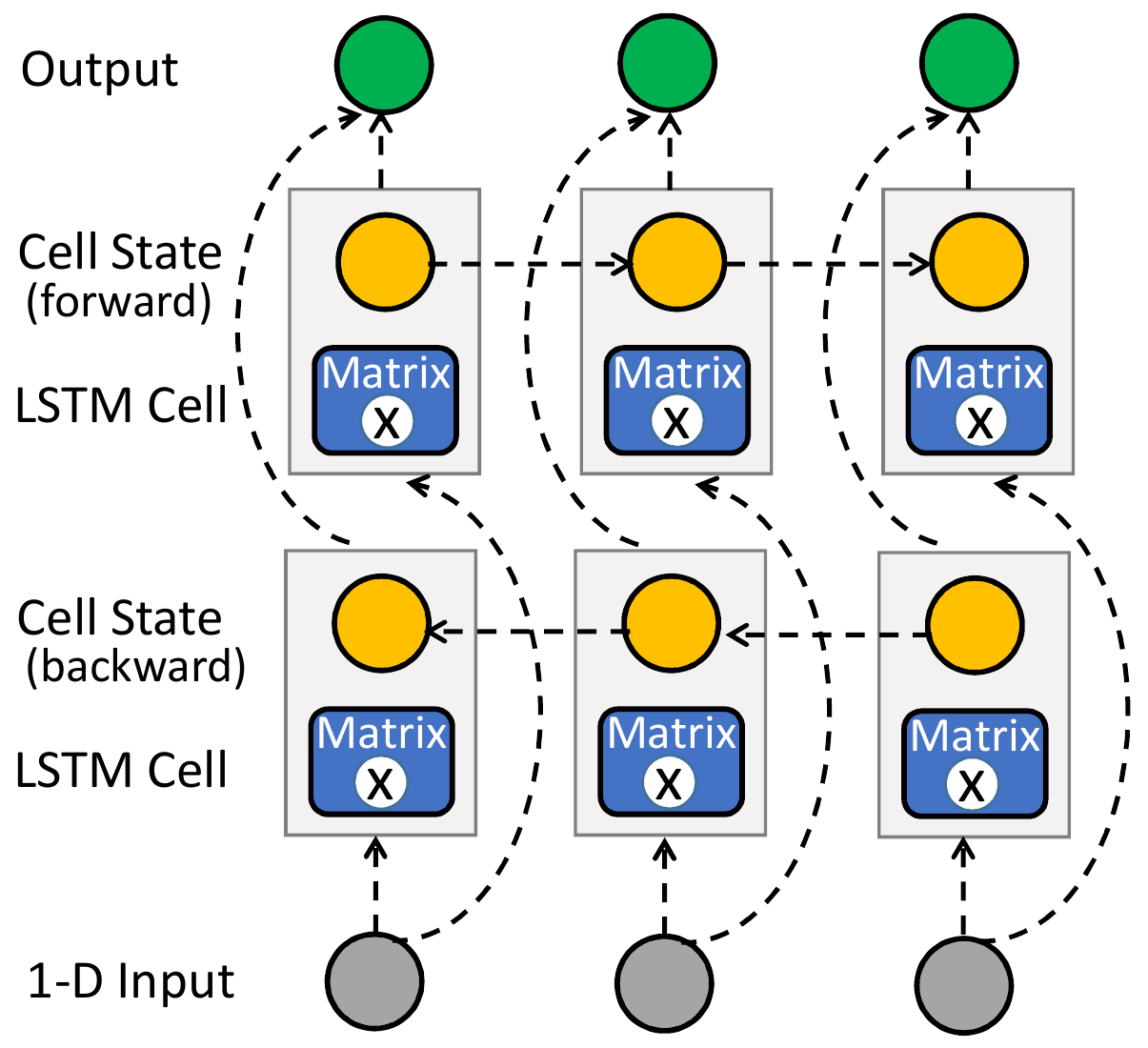}
		\label{fig:bilstm}}
	\subfigure[ConvLSTM]{
		\includegraphics[height=5.4cm]{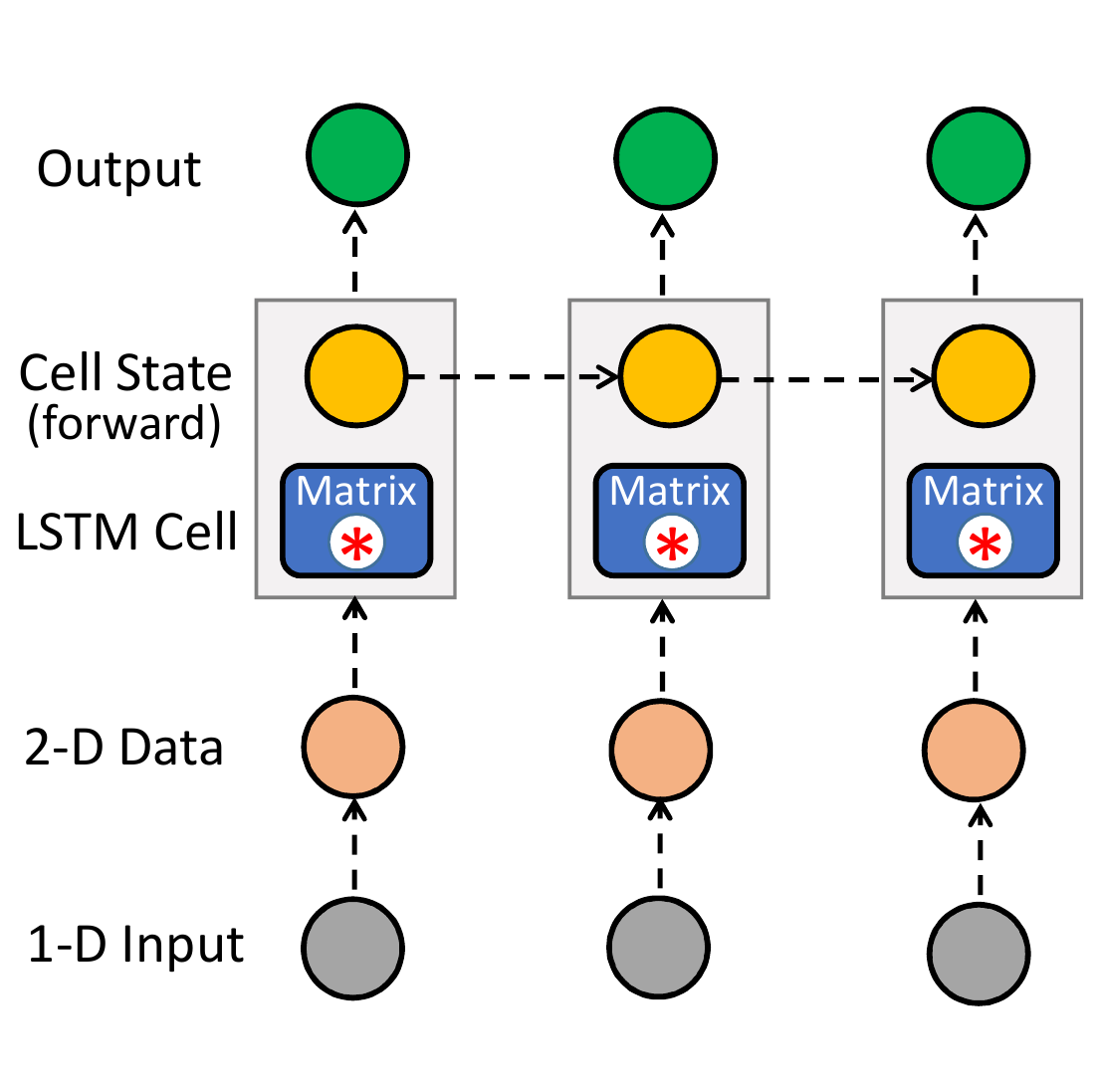}
		\label{fig:convlstm}}
	
	\caption{Network structure comparison of three LSTM variants: (a) UniLSTM, (b) BiLSTM, and (c) ConvLSTM. The symbol $\times$ and $\ast$ represent multiplication and convolution operation, respectively.}
	\label{fig:lstms}
\end{figure*}
\begin{figure*}[]
	\centering
	\includegraphics[scale = 0.6]{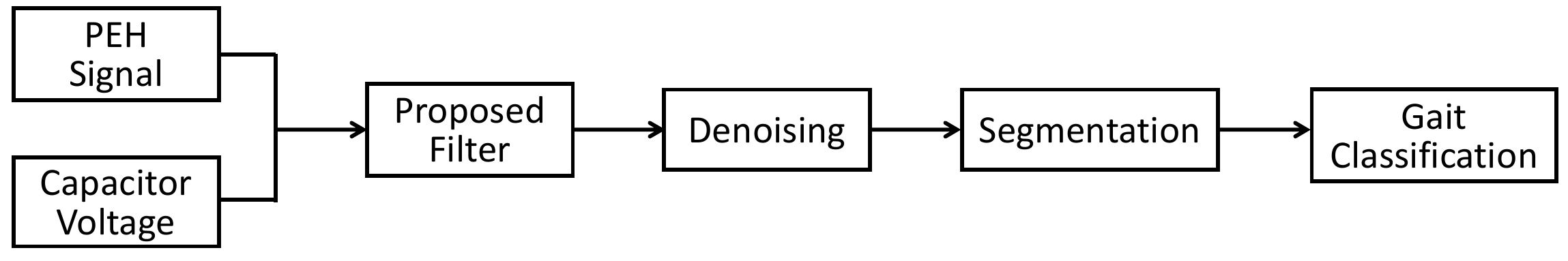}
	\caption{System model of gait recognition using SEHS.}
	\label{fig:system overview}
\end{figure*}

The complexity of the filter is $O(N)$ ($N$ is the number of samples), which suggests that it can be implemented without significant computational burden. Note that the filter requires the capacitor voltage ($V_C$) as input, which can be easily obtained using an additional ADC channel as shown in our circuit design (Figure \ref{fig:circuit}). Use of additional ADC channels contributes minimal power consumption overhead as will be demonstrated in our measurements in Section \ref{section:overhead}.

\section{LSTM-based Gait Classification}
\label{s:gait_recognition}

A gait cycle is basically sequential movements of human body that usually involve six main phases: heel strike, foot flat, mid-stance, heel-off, toe-off, and mid-swing~\cite{shultz2015examination}. These phases are interrelated and the characteristics of former phases interfere with the later ones (i.e., there exists temporal correlation), thereby forming a unique gait pattern of each individual. Thus, we propose to use the long short-term memory (LSTM) neural networks~\cite{hochreiter1997long} for gait classification
due to its superior performance on learning information from
sequences/signals with high temporal correlation~\cite{truong2018capband,zhang2018internet,plotz2018deep}. 
LSTM neural networks exhibit a chain-like structure consisting of multiple cells, where the input of each cell is a sample in a time series data. The core idea behind LSTMs is the cell state which looks like a conveyor belt and preserves the memory of the network~\cite{olah2015understanding}. By dynamically updating the cell state, previous information that is useful to current state can be retained. 

In this work, we consider three variants of LSTM, i.e., unidirectional LSTM (UniLSTM), bidirectional LSTM (BiLSTM)~\cite{graves2005framewise}, and convolutional LSTM (ConvLSTM)~\cite{xingjian2015convolutional}, as compared in Figure~\ref{fig:lstms}. UniLSTM processes sequential data in forward direction thus only preserving information of the past. In contrast, BiLSTM has two cell states which can preserve information of both the past and future by running in forward and backward direction, respectively. In general, BiLSTM has better performance than UniLSTM as temporal correlations from both directions can be learned. ConvLSTM is just like UniLSTM, but internal matrix multiplications are exchanged with convolution operations. It is usually used in sequential image data (e.g., video) processing so that the one-dimensional PEH data should be converted to two-dimensional data before feeding to the LSTM cells.

The three models are designed as follows. Each LSTM variant has one LSTM layer containing $N$ cells, where $N$ is the number of samples in a gait cycle. Each cell is composed of 32 hidden units (the choice of 32 was arrived empirically to prevent over-fitting while keeping good recognition accuracy). On top of that, a fully connected layer is used to convert the output matrix of the LSTM network into a class vector, i.e., the probability of current instance to each class. Then, we apply a \textit{softmax} layer to obtain the final class by selecting the class with maximum probability.

\section{Evaluation}
\label{sec:evaluation}

The system model for evaluating gait recognition using SEHS is illustrated in Figure~\ref{fig:system overview}. Specifically, we first built an insole-based prototype to collect data from subjects. Then, the proposed filter is applied to the collected PEH signal and capacitor voltage to minimize the interference of energy storage. Afterwards, the filtered signal is denoised and segmented before feeding to the LSTM-based classifiers. With the framework, we then evaluate the performance of gait recognition and energy harvesting. Finally, as the filter requires the capacitor voltage as input, we explore its practicability by measuring the power overhead of sampling the capacitor voltage.

\subsection{SEHS Prototyping}
\label{s:proto}
\begin{figure}[]
	\centering
	
	\subfigure[Insole]{
		\includegraphics[scale=0.45]{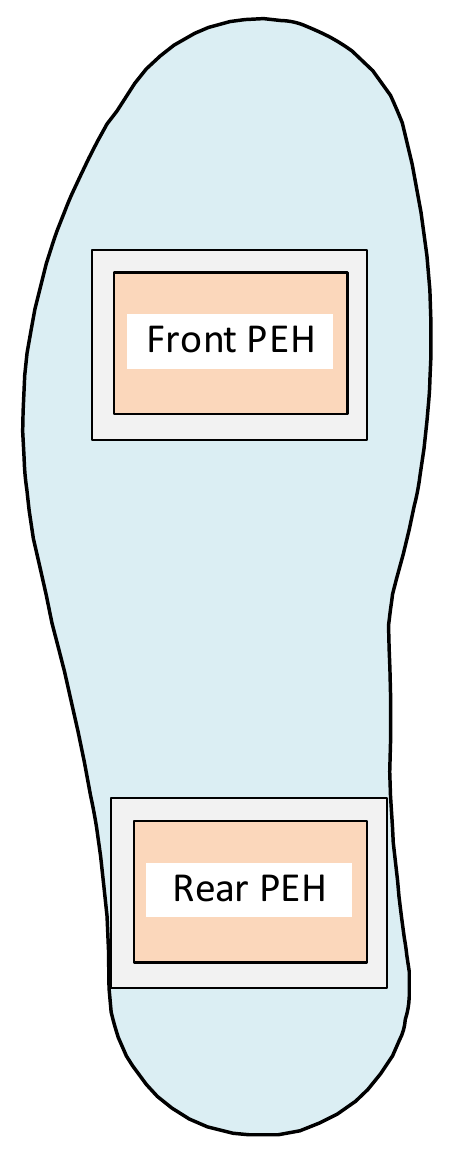}
		\label{fig:insole}}
	\subfigure[Circuit design and placement of the prototype when subject is walking]{
		\includegraphics[scale=0.53]{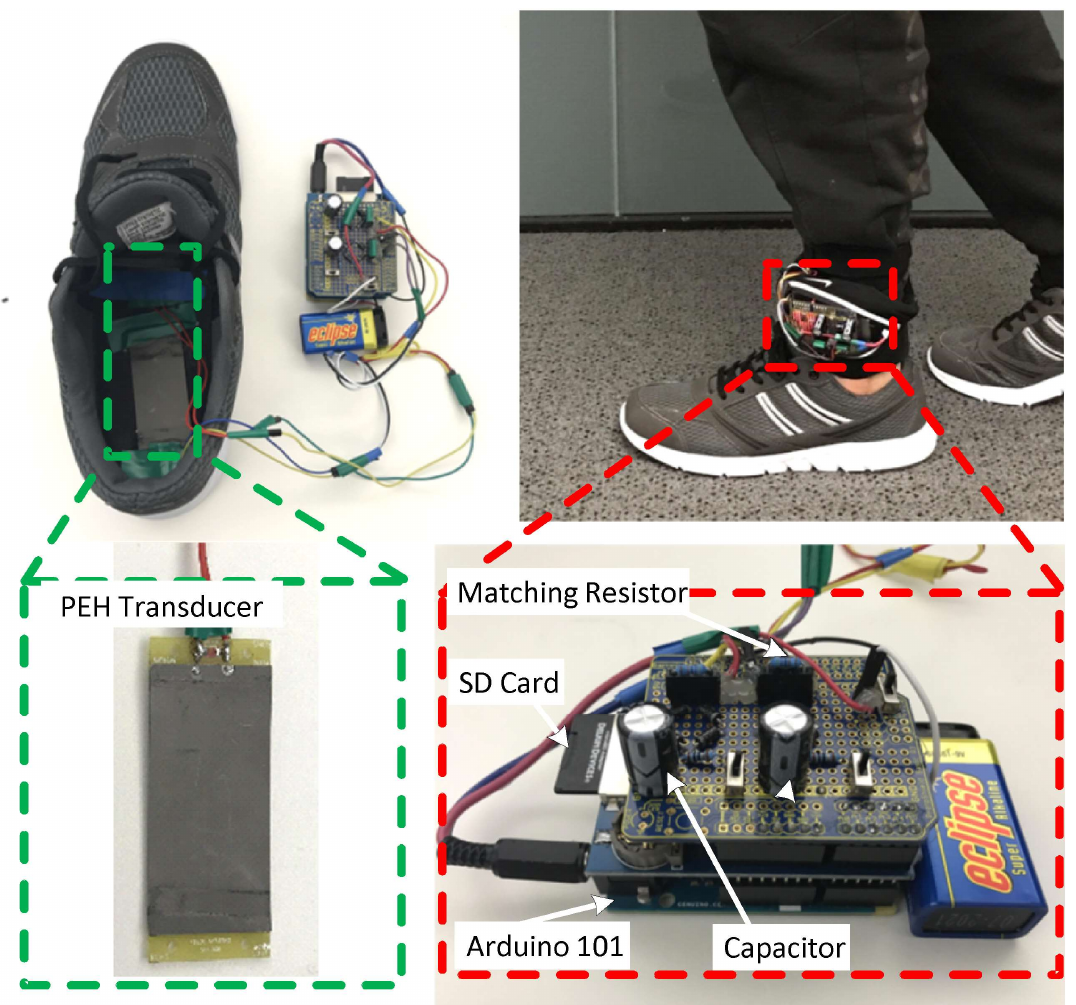}
		\label{fig:apperance}}
	
	\caption{The design and appearance of the SEHS prototype.}
	\label{fig:prototype}
\end{figure}

Figure \ref{fig:prototype} shows the prototype we designed and implemented in the form factor of a shoe to harvest energy during human walking and detect the user gait at the same time using the PEH voltage signal. The prototype includes two PEHs from Piezo System~\cite{piezosystem} mounted on the front and rear of the insole (refers to PEHFront and PEHRear), which are used to investigate the impact of PEH placement on the gait recognition accuracy. AC voltages from the PEHs are rectified by a full-bridge diodes rectifier and charged into two 1000$\mu F$ electrolytic capacitors. The output voltage and capacitor voltage are sampled by an Arduino 101~\cite{arduino101} board,  which is equipped with an Intel Curie microcontroller. A sampling rate of 100 Hz is used for data collection and the sampled data is saved on a 4GB microSD connected to the Arduino using a microSD shield. A nine volts battery powers the whole system. To help users collect data, the prototype contains three switches, one is to control the start and stop of data collection and the other two for controlling the charging and discharging of the two capacitors respectively. The entire Arduino board is placed outside the shoe. 

\begin{figure}[]
	\centering
	\subfigure[Experiment environment]{
		\includegraphics[scale=0.1352]{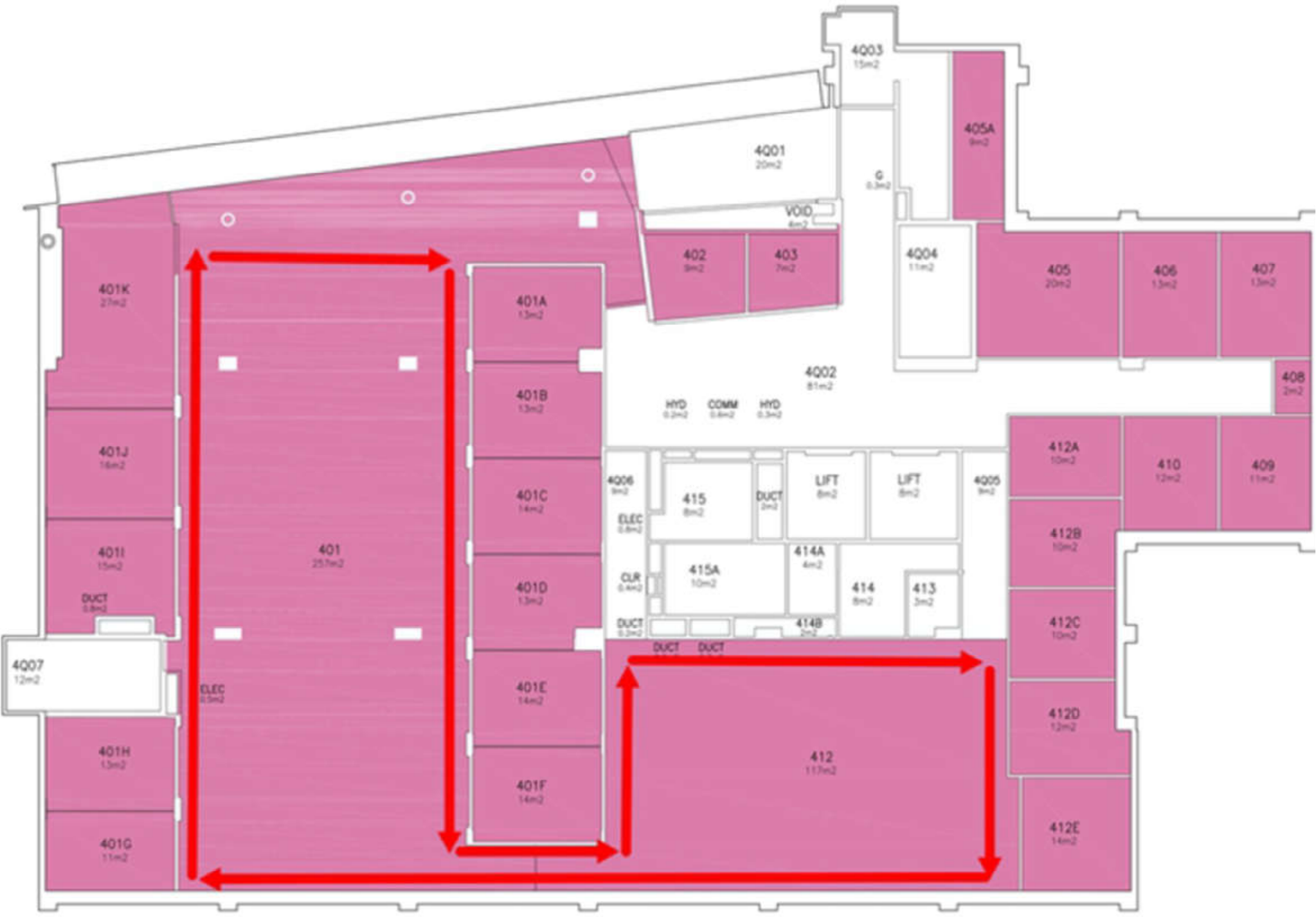}
		\label{fig:route}}
	\subfigure[Subject Walking]{
		\includegraphics[scale=0.36]{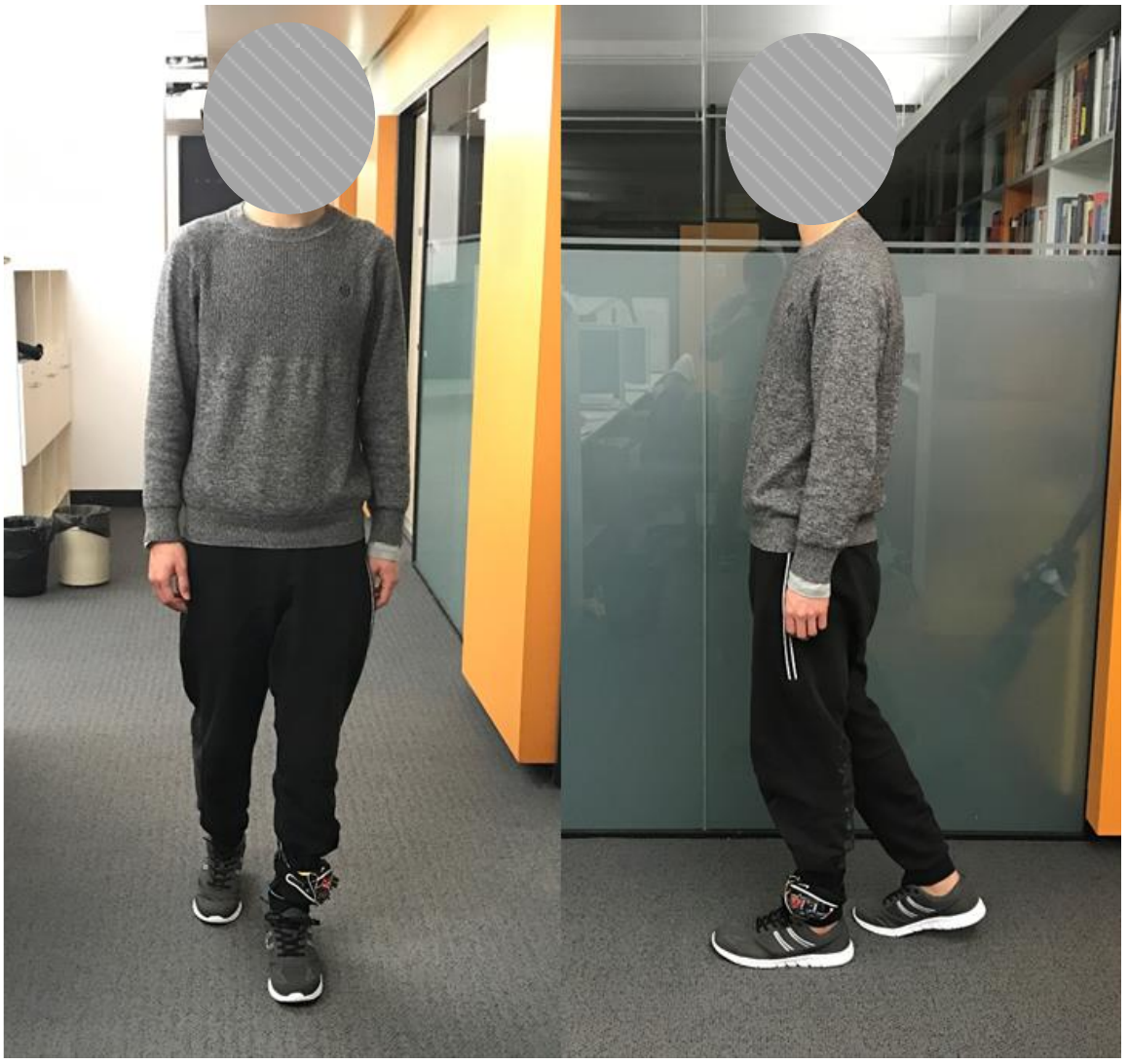}
		\label{fig:volunteer}}
	\caption{Data collection setup.}
	\label{fig:data collection}
\end{figure}

\begin{figure}[]
	\centering
	\includegraphics[scale = 0.38]{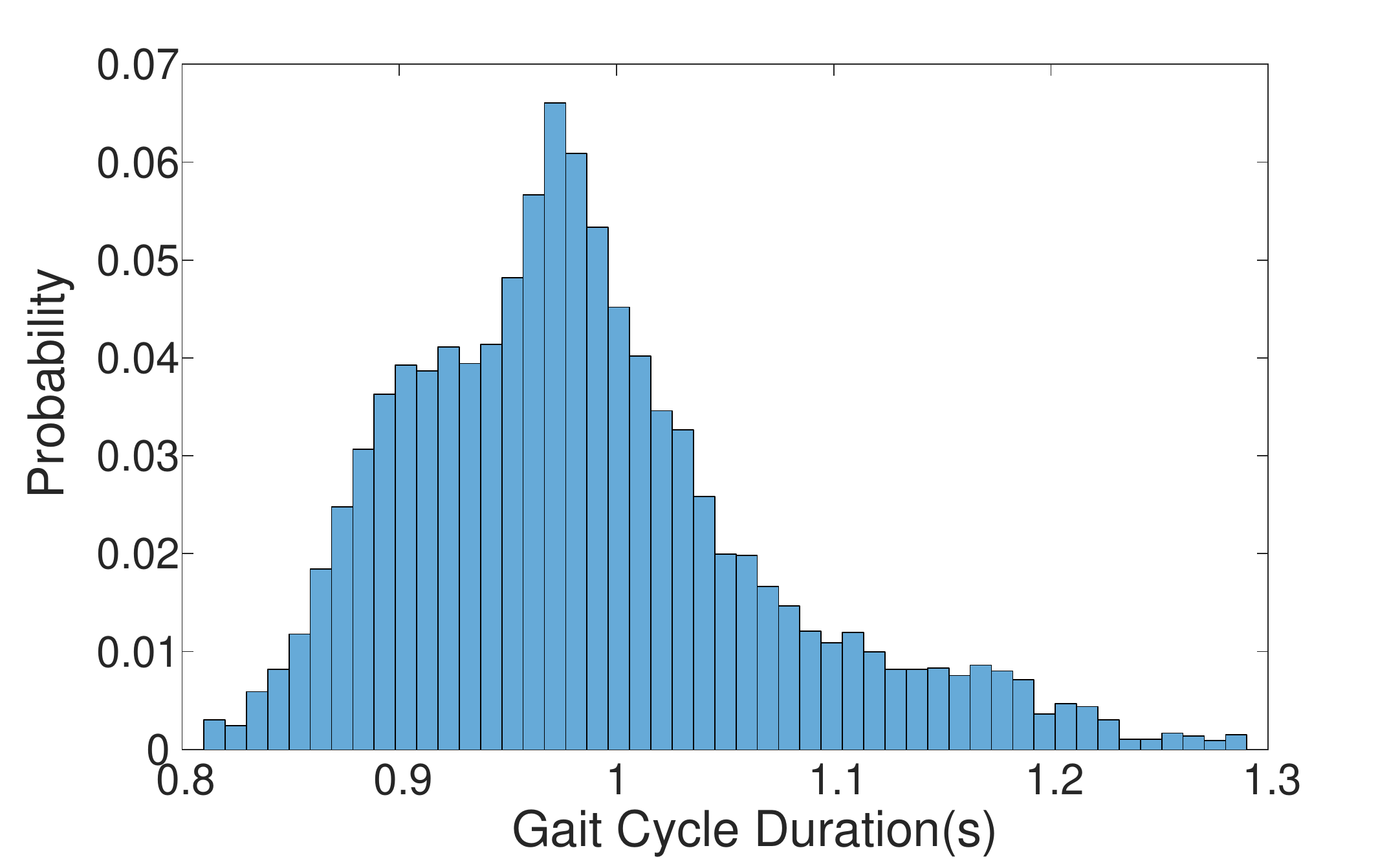}
	\caption{Distribution of gait cycle duration.}
	\label{fig:stepLengthDist}
\end{figure}

The Arduino 101 measures voltage between 0 and 5 volts and provides 10 bits of resolution, i.e., 1024 different values. The corresponding output voltages from the measurements, therefore, are obtained as: 
\begin{equation}
voltage = \frac{5*ADC\;measurement}{1024}
\end{equation}

\begin{figure*}[t]
	\centering
	\begin{minipage}[]{0.66\textwidth}
		\subfigure[]{
			\includegraphics[height=4.44cm]{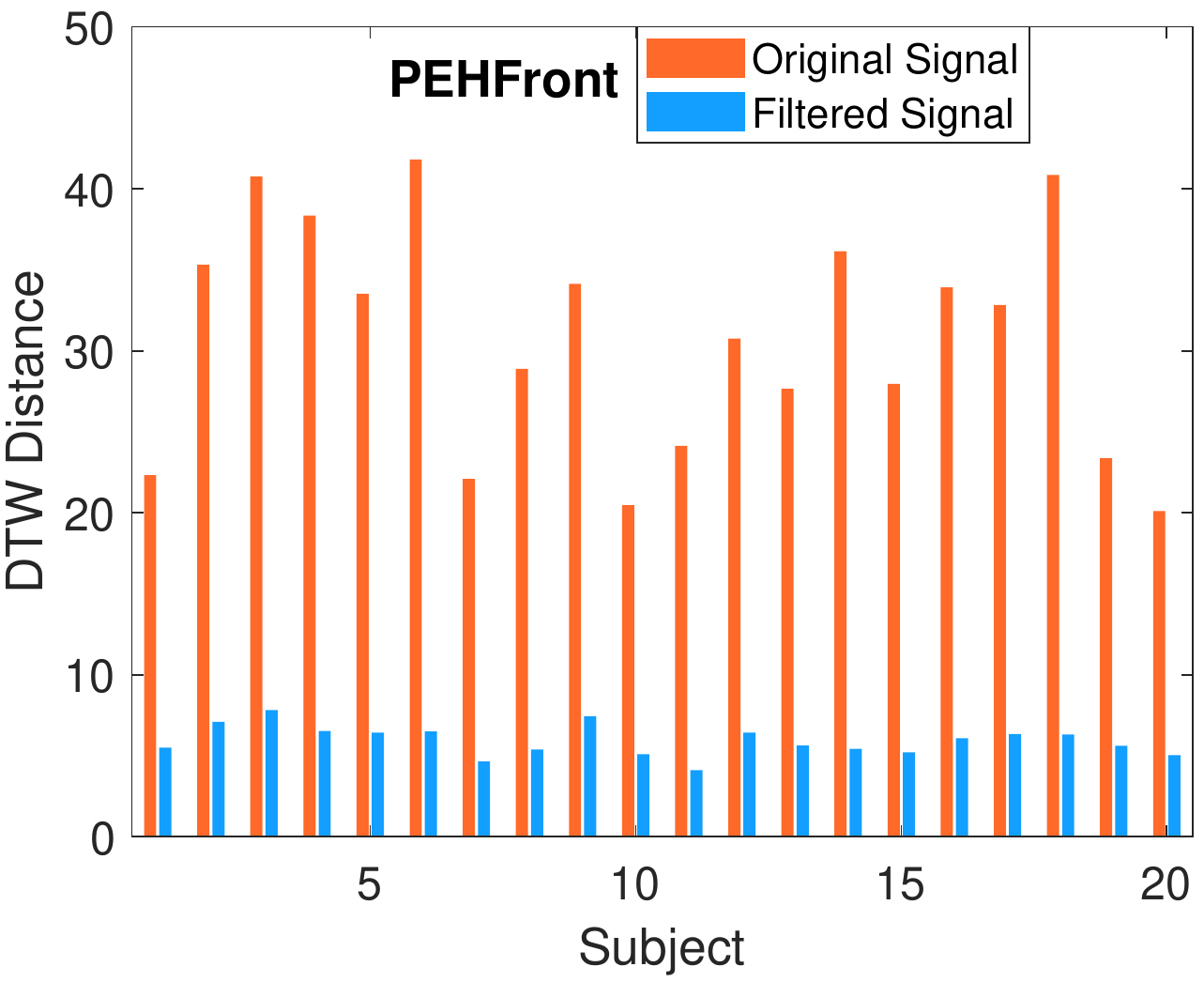}
			\label{fig:dtw_dist_1}}
		\subfigure[]{
			\includegraphics[height=4.44cm]{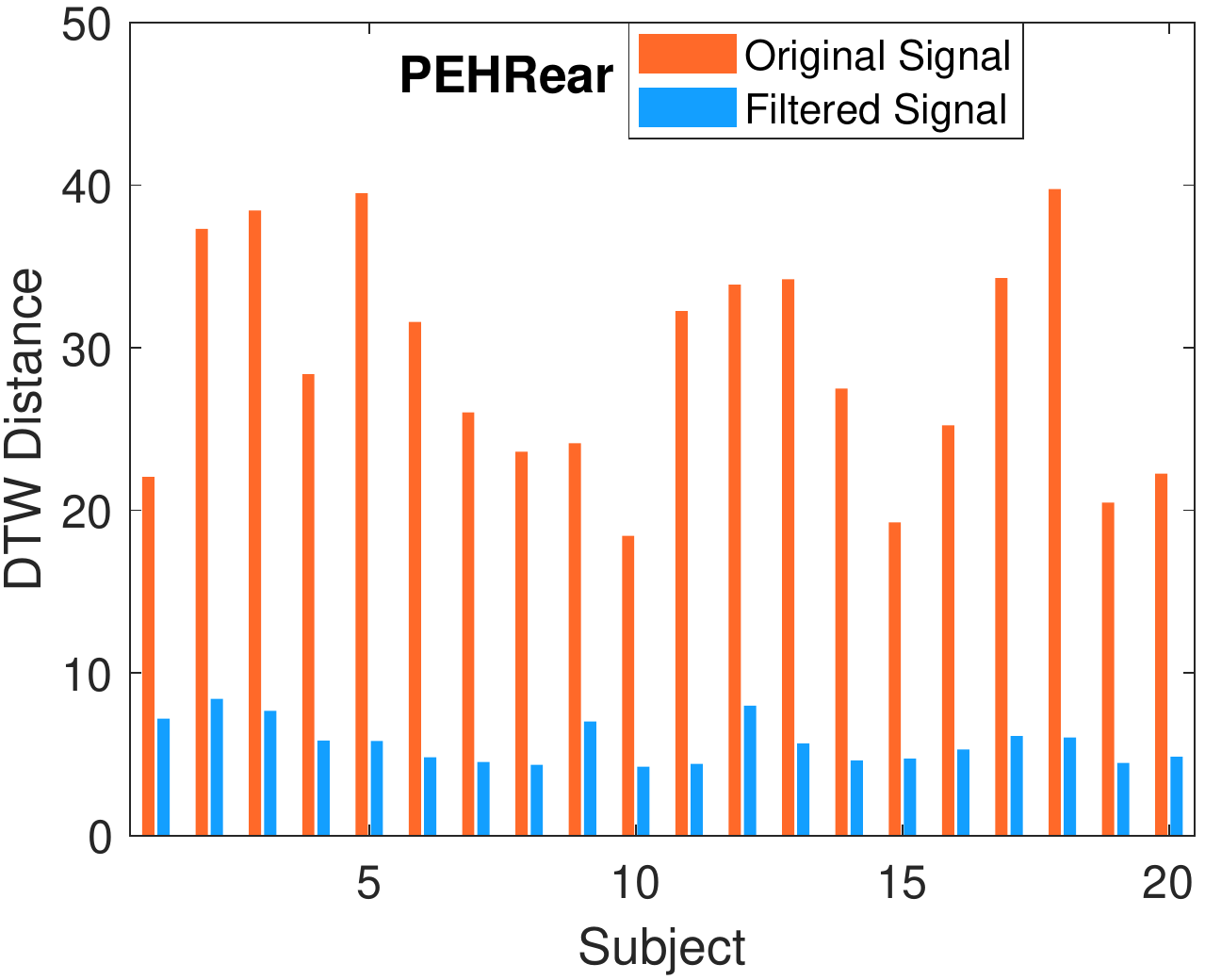}
			\label{fig:dtw_dist_2}}
		\caption{Similarities among individual gaits of the same subject for original vs. filtered signals: (a) front PEH and (b) rear PEH. Lower DTW distance means higher similarity.}
	\end{minipage}
	\begin{minipage}[]{0.33\textwidth}
		\includegraphics[height=5.14cm,width=6.3cm]{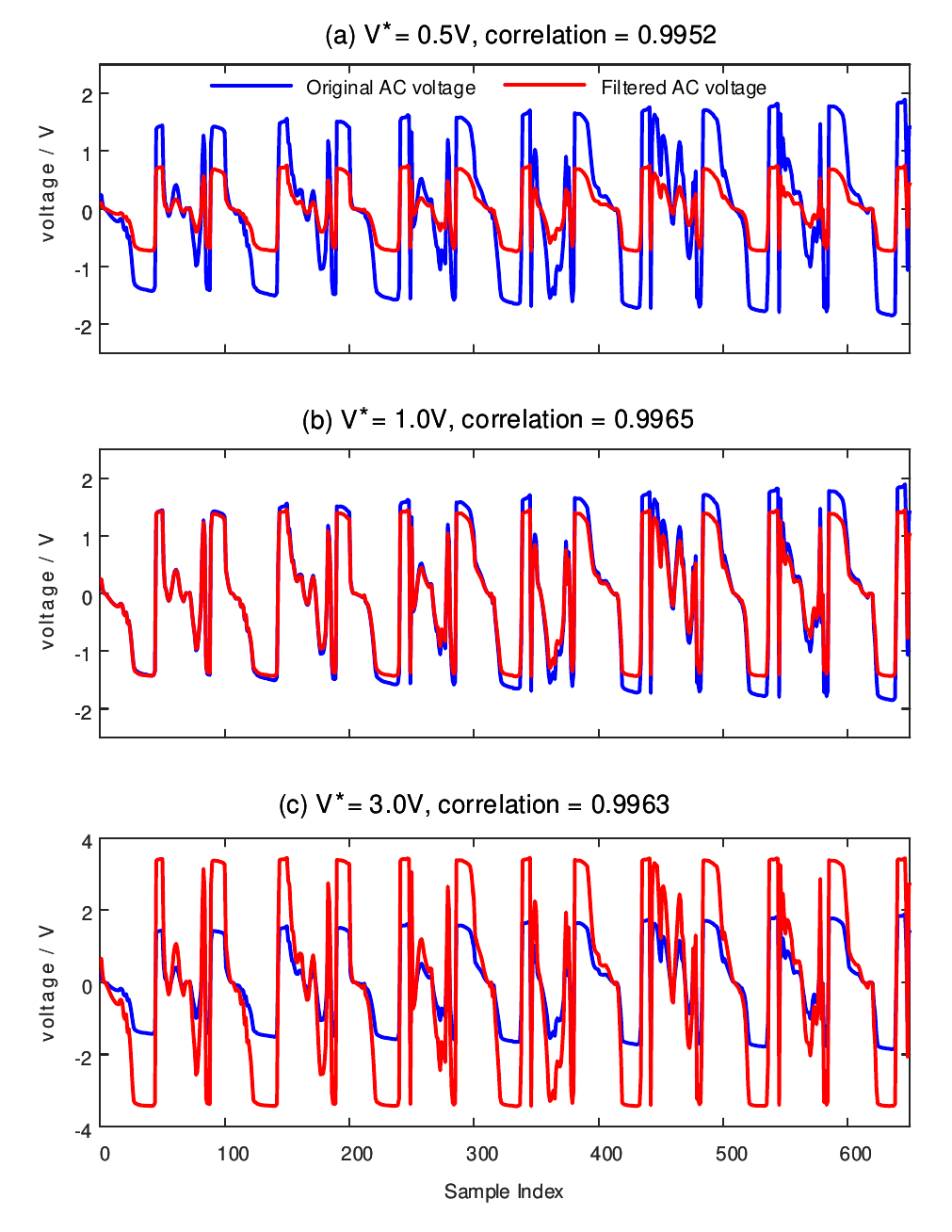}
		
		\caption{Waveform comparison for AC voltage with and without filtering.}
		\label{fig:normalization}
	\end{minipage}
\end{figure*}

\subsection{Data Collection and Pre-processing}
During the data acquisition stage, we create the dataset by asking 20 healthy volunteers, including 16 males and 4 females\footnote{Ethical approval for carrying out this experiment has been granted by the corresponding organization (Approval Number HC15888).}, to walk along the specified route as shown in Figure~\ref{fig:route}. Each volunteer is asked to wear a shoe equipped with the designed insole in their left foot and the data collection module is attached to the ankle position as shown in Figure~\ref{fig:volunteer}. The participants are suggested to walk in his/her normal walking style and speed. The data collection duration is approximately 300 seconds and the subjects need to toggle the switch to discharge the capacitors every 50 seconds. The reason is that the accumulated voltage on the capacitor will exceed 5$V$ when walking for a certain period of time (around 55 seconds in our tests), so that it can not be measured. More importantly, high voltage may damage the ADC module of the board. 

We collect a time series of voltage signal of the PEHs when subjects are walking, in which the signal follows a cyclic pattern reflecting the gait of each subject. As shown in Figure~\ref{fig:rawVolcompare}(b), there are two peaks within one gait cycle, which indicate the heel strike and toe-off time of the foot respectively. Firstly, the moving average function and a low pass filter with the cutoff frequency of 10 Hz are used to eliminate out-band interference. We then apply a bandpass filter to detect these peaks and the gait cycle is obtained by combining the samples between two consecutive peaks together. Since the walking pattern varies from different subjects, we tune the lower and upper cutoff frequency of the bandpass filter for each subject ranging from 0.5Hz to 3Hz to enable an accurate gait cycle segmentation. 

After the peak detection and samples combination, gait cycles with different time duration are available. Figure~\ref{fig:stepLengthDist} plots the distribution of the gait cycle duration of the 20 subjects who are suggested to walk in their normal speed. It is clear that the time duration of one gait cycle varies from 0.8s to 1.3s. To deal with such variable walking speeds that may occur among different subjects or one subject in different walking scenarios, we perform the linear interpolation on the detected gait cycles. According to the distribution of the gait cycle length, the number of samples in one gait cycle ranges from 80 to 130 with the sampling rate of 100 Hz. Thus, we interpolate the gait cycles to equal length with 130 samples.

Accurate gait pattern extraction is a critical factor that affects gait recognition accuracy. In our experiment, volunteers were asked to walk in a square environment as shown in Figure~\ref{fig:route} for several minutes, during which they experienced several turnings and some short pauses. Obviously, the detected gait cycles within these periods contain distorted gait patterns. Therefore, it is important to omit these unusual cycles to keep high recognition accuracy. We employ Dynamic Time Warping (DTW) to delete unusual cycles for each subject. In detail, the average distance of all the gait cycles are computed firstly and treated as the typical cycle. Then, the distance between each cycle and the typical cycle is calculated. We detect and omit the irregular cycles by a simple threshold method, i.e. if the DTW distance of a detected cycle is higher than a predefined value, it will be dropped. The DTW distance reflects the similarity among a given cycle and the rest cycles of the subject. We collect around 280 to 300 gait cycles for each subject. To achieve a fair classification, we extract 250 gait cycles per subject after performing irregular cycle deletion. In total, a 2x20x250 gait cycle dataset is created and utilized to evaluate the performance of the proposed SEHS architecture and filter.

\subsection{Training and Classification}

Then, we use the three designed LSTM networks for gait classification. We split the collected dataset into the training set (80\%) and testing set (20\%), and train the models with 5-fold cross validation mechanism. During training, the loss function and optimizer are set to cross-entropy and Adam, respectively. In addition, we set batch size to 64 and utilize the \textit{EarlyStopping}~\cite{prechelt1998automatic} mechanism to prevent model over-fitting. After training, the testing set is delivered to the trained models and the classification results are obtained for accuracy calculation. All the procedures are implemented with Keras in Python running on Tensorflow 2.0 platform.

In addition, we consider three benchmark classifiers: Support Vector Machine (SVM), K-Nearest Neighbor (KNN), and
Sparse Representation based Classification (SRC)~\cite{huang2007sparse}. SVM and KNN are two typical machine learning based classifiers adopted in a multitude of sensing applications like human activity recognition. We extract
22 statistical features~\cite{khalifa2018harke} from both time and frequency domain for each
gait cycle and use them as the input to the two classifiers. SRC automatically mines features using sparse representation and has been proved to be more robust to environmental noise in sensing tasks. Moreover, it was applied to the state-of-the-art~\cite{xuTMCGait} that uses PEH for gait recognition. Like LSTM, SRC can be regarded as a feature-less classifier so that we feed complete gait cycles to it during classification. We perform 5-fold cross-validation on the collected dataset and all the classifications for the three benchmarks are carried out in MATLAB.

\subsection{Performance Evaluation}

Next, we evaluate the performance of the proposed SEHS architecture, filter, and LSTM neural networks based on the collected dataset in the following aspects. Firstly, we use DTW distance~\cite{berndt1994using} and Pearson correlation~\cite{benesty2009pearson} to explore the effectiveness of the proposed filtering algorithm from the signal waveform aspect. Secondly, we evaluate the filter from the perspective of gait recognition performance (e.g., precision and recall), under different system parameters, such as sampling rate and classifier.

\subsubsection{Filter Performance}

Since gait cycles of the same person are expected to be similar during a normal walk, such \textit{gait similarity} can be used to assess the performance of the proposed filter algorithm. Due to the distortion effect of energy harvesting, we can expect that the gait similarity among different cycles for the original AC signal would be low, but would increase when the filter is applied. For each subject and each PEH, we computed gait similarity for a given trace of 250 gait cycles as the average DTW (dynamic time warping) distance between all possible pairs of gait cycles. The lower the DTW distance, the higher the gait similarity for samples from each subject. Figures~\ref{fig:dtw_dist_1} and \ref{fig:dtw_dist_2} compare DTW distance of signals from the two PEHs for each of the 20 subjects. We can clearly see that the gait similarities for the filtered signal are consistently higher (or the DTW distance is 5$\times$ lower) than those for the original signal, irrespective of the subjects and position of PEH. This provides evidence that the proposed filtering algorithm has successfully reduced signal distortions caused by energy harvesting.

\begin{figure*}[]
	\centering
	\begin{minipage}[]{0.495\textwidth}
		\hspace{0.1in}
		\includegraphics[scale=0.55]{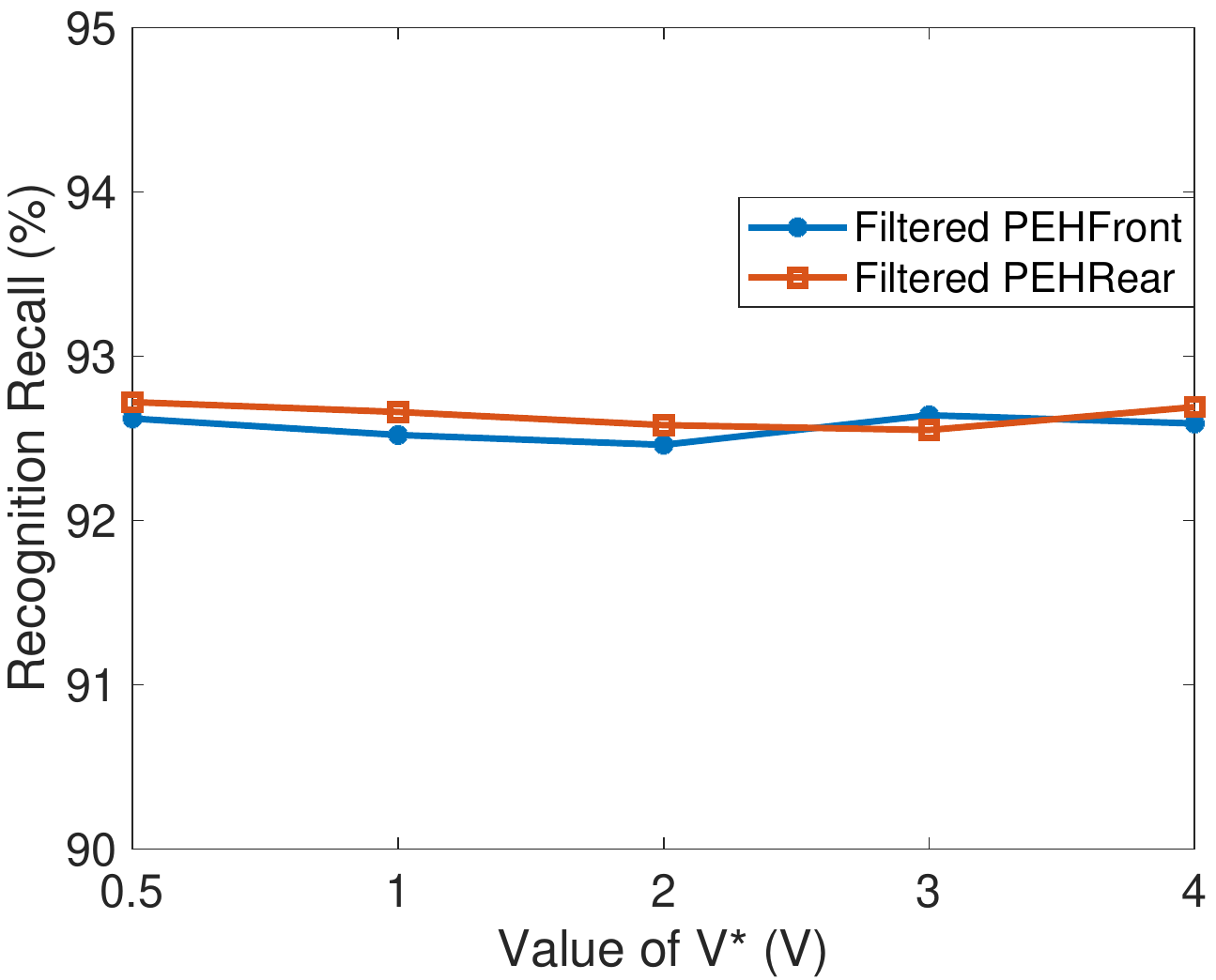}	
		\caption{Recall vs. Value of $V^*$. }
		\label{fig:V}
	\end{minipage}
	\begin{minipage}[]{0.495\textwidth}
		\hspace{0.1in}
		\includegraphics[scale=0.55]{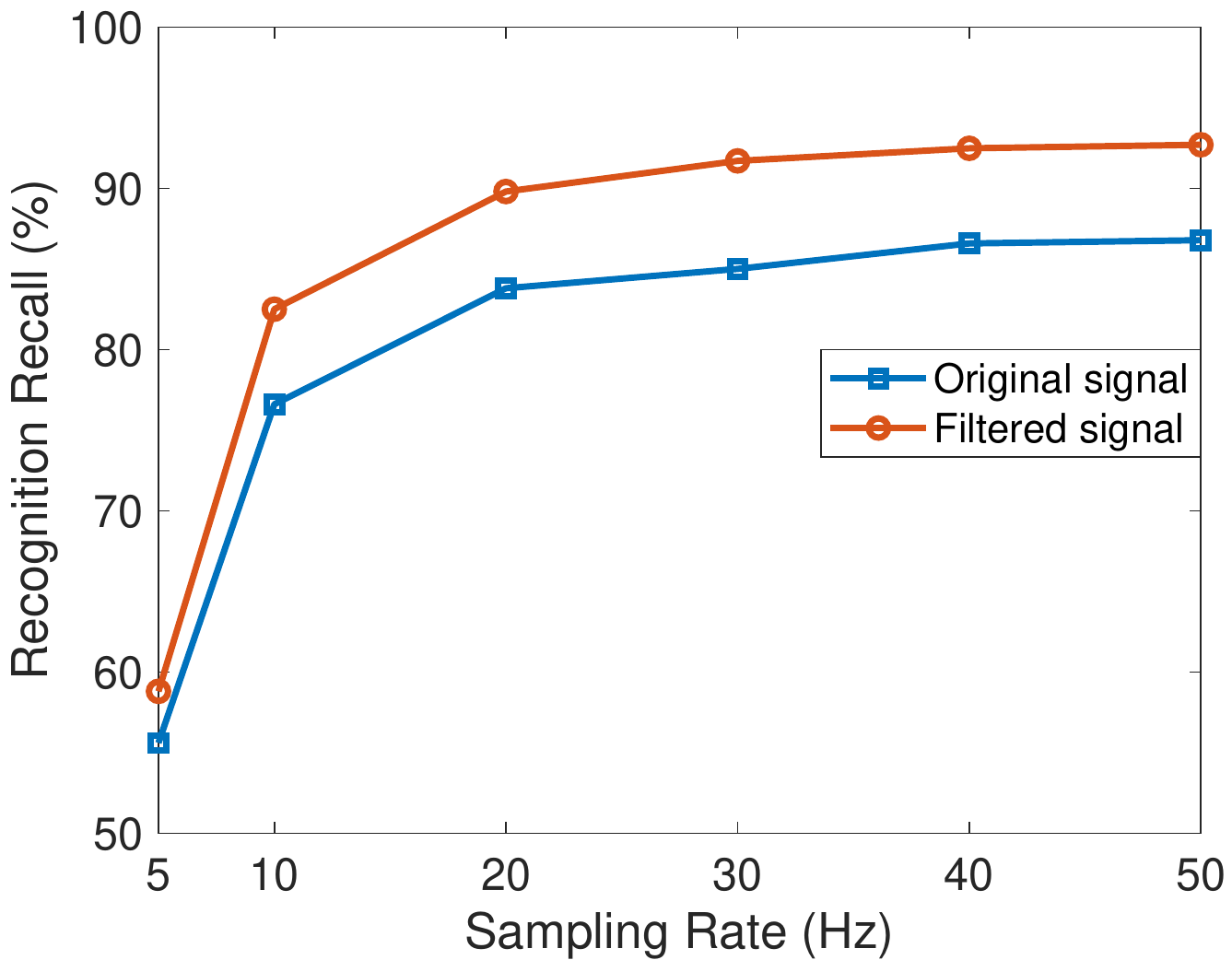}		
		\caption{Recall vs. Sampling rate.}
		\label{fig:accuracySampling}
	\end{minipage}
\end{figure*}

\begin{table*}[]
	\centering
	\caption{Recognition performance vs. classifiers.}
	\label{tab:Classifiers}
	\ra{1.3}
	\setlength\tabcolsep{2.25pt}
	\begin{tabular}{lccccccccc}\toprule
		\multirow{4}{*}{\textbf{Classifier}}& \multicolumn{4}{c}{\textbf{Original signals }} & \phantom{a}& \multicolumn{4}{c}{\textbf{Filtered signals }} \\
		\cmidrule{2-5} \cmidrule{7-10} 
		& \multicolumn{2}{c}{\textbf{PEHFront}} &\multicolumn{2}{c}{\textbf{PEHRear}} &&\multicolumn{2}{c}{\textbf{PEHFront}} & \multicolumn{2}{c}{\textbf{PEHRear}} \\ \cmidrule{2-5} \cmidrule{7-10}
		
		& \textbf{Precision} & \textbf{Recall}  & \textbf{Precision} & \textbf{Recall}   && \textbf{Precision} & \textbf{Recall}  & \textbf{Precision} & \textbf{Recall} 
		
		\\ \midrule
		KNN & 55.43 ($\pm$0.64)\% & 54.77 ($\pm$0.47)\%& 51.87 ($\pm$0.38)\% & 50.99 ($\pm$0.35)\% &&63.75 ($\pm$0.55)\% &62.89 ($\pm$0.60)\% & 62.03 ($\pm$0.45)\%&60.54 ($\pm$0.50)\% \\
		
		SVM & 67.82 ($\pm$0.73)\% & 66.76 ($\pm$0.68)\%& 66.68 ($\pm$0.63)\% & 65.79 ($\pm$0.65)\% &&71.47 ($\pm$0.26)\% &74.28 ($\pm$0.29)\% & 73.85 ($\pm$0.53)\%& 72.90 ($\pm$0.48)\% \\
		SRC &  77.19 ($\pm$0.47)\% & 76.56 ($\pm$0.51)\%& 76.57 ($\pm$0.52)\% & 76.18 ($\pm$0.57)\% &&85.86 ($\pm$0.62)\% &85.42 ($\pm$0.58)\% & 87.02 ($\pm$0.39)\%& 86.67 ($\pm$0.42)\% \\
		ConvLSTM & 85.26 ($\pm$0.53)\% & 84.87 ($\pm$0.55)\%& 84.56 ($\pm$0.55)\% & 83.95 ($\pm$0.62)\% &&90.85 ($\pm$0.26)\% & 90.57 ($\pm$0.25)\% & 91.63 ($\pm$0.18)\%& 91.46 ($\pm$0.27)\% \\
		UniLSTM & 86.87 ($\pm$0.54)\% & 86.50 ($\pm$0.58)\%& 85.62 ($\pm$0.43)\% & 85.30 ($\pm$0.44)\% &&91.63 ($\pm$0.33)\% & 91.50 ($\pm$0.35)\% & 92.51 ($\pm$0.35)\%& 92.41 ($\pm$0.36)\% \\

		\textbf{BiLSTM} & \textbf{96.79 ($\pm$0.37)\%} & \textbf{96.77 ($\pm$0.38)\%}& \textbf{96.98 ($\pm$0.59)\%} & \textbf{96.89 ($\pm$0.59)\%} && \textbf{98.41 ($\pm$0.49)\%} & \textbf{98.36 ($\pm$0.51)\%} & \textbf{98.95 ($\pm$0.40)\%}& \textbf{98.94 ($\pm$0.41)\%} \\
		\bottomrule
	\end{tabular}
\end{table*}

Figure~\ref{fig:normalization} compares the waveform of the original AC voltage (blue)
	and the filtered version (red) for different values of $V^*$. Visually, we can see that irrespective of the value chosen for $V^*$, the filtered signal matches well with the original signal. In addition, we calculate the Pearson correlation coefficient 
	between the original signal and filtered signal. The Pearson correlation coefficient ranges between $-1$ and 1, where 1 means total positive linear correlation, 0 means no linear correlation, and $-1$ means total negative linear correlation. For different $V^*$, the calculated correlation coefficients are all around 0.996, suggesting that our filter successfully retains the gait pattern of the original signal. Next, we investigate whether the value of $V^*$ will affect the gait recognition performance.

\subsubsection{Gait Recognition Performance}
\label{s:filter performance}
\textbf{\qquad  Impact of $V^{*}$: }
To do that, we applied the proposed filter with different value of $V^{*}$ (from 0.5V to 4V) on the original signal and calculate the recognition recall of the filtered signals. All the three LSTM variants are considered and the PEHFront and PEHRear are evaluated separately. As shown in Figure~\ref{fig:V}, all the curves are almost flat, indicating that recognition recall is not affected by the chosen of $V^*$. So, we select a typical value of 2V for the rest of evaluation.

\textbf{Gait Recognition Recall vs. Sampling Rates: }
\label{section:sampling}
Given that walking is a low-frequency activity, a sampling rate of 100 Hz might be excessive for capturing the latent gait patterns. We downsample each gait cycle to investigate the minimal sampling rate required, which would result in less sensing and computation power consumption. For different sampling rates, Figure~\ref{fig:accuracySampling} compares the gait recognition recall obtained with and without filtering for both PEHs using the three LSTM-based classifiers. 
	It is clear that for all sampling rates, filtered signal outperforms the original counterpart. In addition, the recognition recall starts to saturate when sampling rate is higher than 40 Hz, which suggests that such sampling rate is sufficient to capture the gait patterns. We will therefore consider a sampling rate of 40 Hz in the subsequent analysis.



\textbf{Gait Recognition Recall vs. Classifiers: }
\label{section:comparealgorithm}
Then, we evaluate the gait recognition performance of the filter against different classifiers. The parameters in SVM and KNN are tuned to provide the highest accuracy. For SVM classifier, we choose quadratic kernel and the box constraint level is set to 5. For KNN classifier, the number of nearest neighbours is set to 10.  For each classifier, we train five different models while do inference with a fixed testing dataset. The average recognition accuracy (precision and recall) and corresponding standard deviation are presented.

Table~\ref{tab:Classifiers} compares the gait recognition performance of the proposed three LSTM variants and three benchmark classifiers. We can clearly see that the filtered signals outperform the original counterparts for all the six classifiers in terms of recognition precision and recall. Our filter improves the recognition recall by 8\% - 10\% for machine learning based classifiers and 2\% - 7\% for LSTM based classifiers, which provides strong evidence that the proposed filter is robust and effective across the choice of classifiers. This result also reveals that LSTMs show better performance in learning from noisy PEH data, while conventional machine learning methods struggle without filtering.

For the three LSTM variants, BiLSTM achieves the best performance due to its capability to learn temporal correlation from both the forward and backward direction. Compared to UniLSTM, ConvLSTM achieves even lower accuracy which might because the temporal correlation is deteriorated when converting the 1-D sequential signal into 2-D data. In addition, compared to previous work~\cite{xuTMCGait} using SRC for PEH-based gait recognition, the proposed BiLSTM classifier significantly improves the recognition recall by up to 12\%. Another interesting finding is that the front PEH and rear PEH achieve almost identical accuracy (within error margin), which suggests that both PEH are capable of capturing gait information of the wearer. 

\begin{figure}
	\centering
	\includegraphics[scale=0.55]{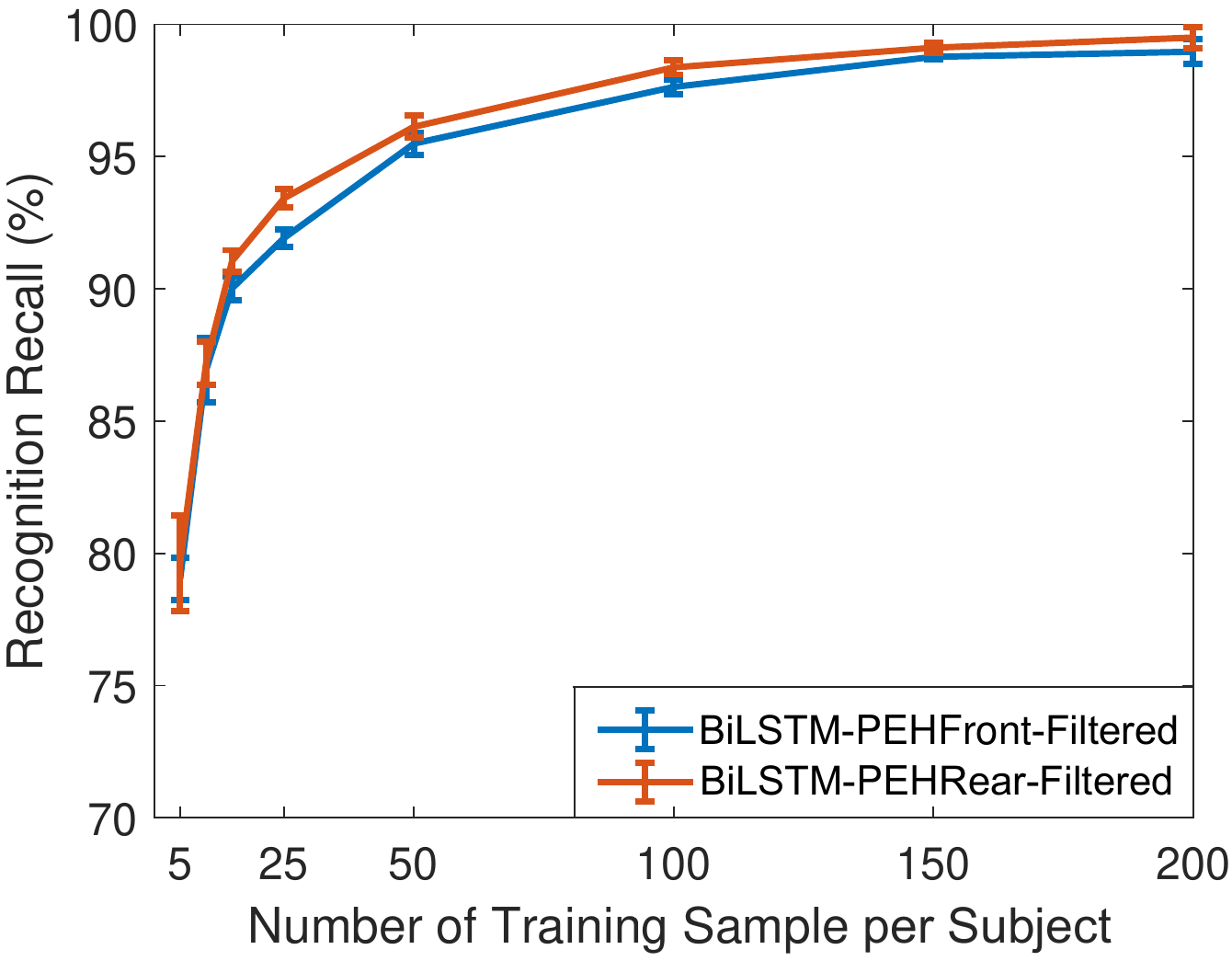}		
	\caption{Recall vs. Training size. }
	\label{fig:samplesize}
\end{figure}

\textbf{Gait Recognition Recall vs. Training Sample Size: }
In the above analysis, we perform 5-fold cross-validation on the collected dataset, i.e., 200 gait cycles from each subject are used to train the model. On one hand, collecting massive amount of data increases the burden on users. On the other hand, given that deep learning usually requires abundant training data, it is unsure whether the collected dataset is large enough to fully learn the latent gait patterns. So, with filtered data from both PEHs, we fix the testing dataset and reduce the number of training sample to investigate how the recognition recall changes. The experiment is run with BiLSTM due to its superior performance. As shown in Figure~\ref{fig:samplesize}, the recognition recall grows sharply with the increasing training size when the number of training sample is small, while it starts to saturate when each subject provides 150 samples. This reveals that the collected dataset is enough for the proposed shallow deep neural networks (one layer). We also observe that with 50 training samples per user, a recognition recall of 96\% is obtained, which achieves a balance between user burden and recognition performance.

\subsection{Analysis of Harvested Energy}
In SEHS, not only the voltage signal is collected, but also the generated energy by human walking is stored in the two capacitors. By measuring the capacitor voltage $V$, we can calculate the amount of generated energy from different PEHs and subjects using $E=\frac{1}{2}CV^2$, where $C$ is the capacitance.

\begin{figure}[]
	\centering
	\includegraphics[scale = 0.68]{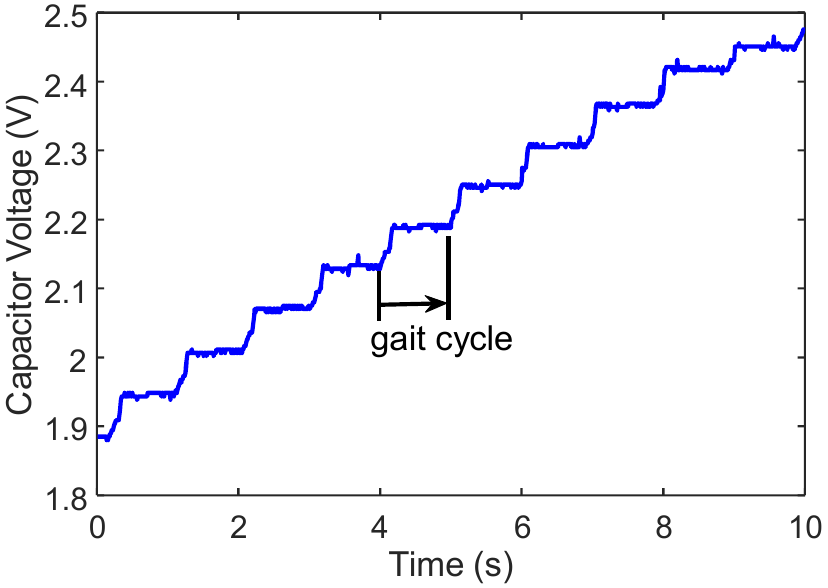}
	\caption{Capacitor voltage during walking. }
	\label{fig:capacitorVol}
\end{figure}

\begin{figure}[t]
	\centering
	\includegraphics[scale = 0.44]{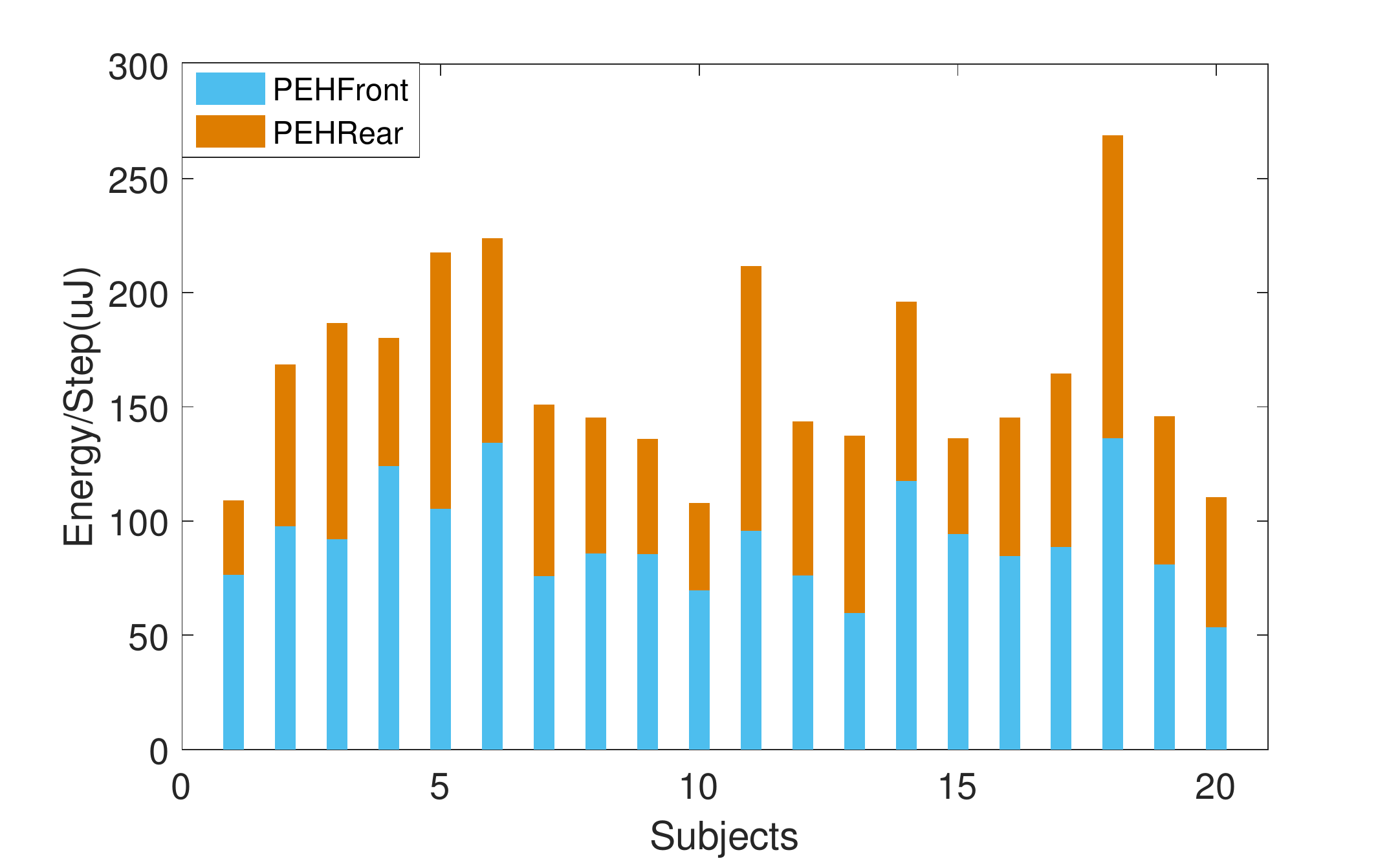}
	\caption{The amount of energy generated per step from the two PEHs. The average generated energy of \textit{PEHFront}, \textit{PEHRear} and total is around 92$uJ$, 72$uJ$ and 164$uJ$ respectively.}
	\label{fig:energyCompTotal}
	\vspace{-0.2in}
\end{figure}

Figure~\ref{fig:capacitorVol} presents the capacitor voltage of one subject, in which each \textit{stair} corresponds to one gait cycle. The \textit{stair} suggests that the energy is only produced within a small time slot, where the capacitor voltage climbs sharply, during each gait cycle. The stair-like capacitor voltage can be utilized for step counting as well~\cite{lan2017capsense}. The distribution of the average generated energy per step of the 20 subjects from the two PEHs is shown in Figure~\ref{fig:energyCompTotal}. It is apparent that the total amount of the harvested energy varies with different people (due to weight and walking style) ranging from $109 \mu J$/step to $269\mu J$/step with an average of $164\mu J$/step ($92\mu J$ for PEHFront and $72\mu J$ for PEHRear). Assume that people walk in 2Hz, i.e., one gait cycle each second, a power output of 164$\mu W$ can be achieved by wearing the insole in one foot, which is much more than the previous work~\cite{xuTMCGait} with only 1$\mu W$ power by harvesting energy from walking-induced vibrations. More practically, such power generation is encouraging to extend the battery lifetime or even replace the battery for wearable devices. 

Due to the realization of simultaneous energy harvesting and sensing, the proposed architecture can increase the amount of harvested energy as well. Specifically, when two PEHs are mounted on the insole, SEHS can harvest energy with both PEHs (i.e., $72+92=164\mu J$) while the separated PEH architecture~\cite{xiang2013powering} has to choose one PEH for sensing and the other for energy harvesting. Around 78\%\footnote{$(164-92)/92=78\%$.} more energy can be harvested in SHES if the front PEH is selected for energy harvesting and the improvement further lifts to 127\%\footnote{$(164-72)/72=127\%$.} when using the rear PEH for energy harvesting.

\vspace{-0.1in}
\subsection{Power Consumption Measurements}
\label{section:overhead}
To this end, we have demonstrated the superior performance of the proposed SEHS architecture in terms of both context detection and energy harvesting. However, the filter in our prototype requires the capacitor voltage samples as an input, which is not used in the previous architectures~\cite{xiang2013powering,xuTMCGait, lan2019entrans}. Consequently, the power consumption overhead of sampling the capacitor voltage should be measured to assess the practicability of the filter. In fact, for each PEH, the proposed SEHS architecture uses a total of three ADCs (two for AC voltage and one for capacitor voltage), while previous method~\cite{xiang2013powering} uses only one ADC channel but with an amplifier consuming around 500$\mu W$ power. Thus, we compare the overall sensing power consumption by measuring the power consumption of sampling ADCs.

\vspace{-0.1in}
\subsubsection{Measurement setup} 

\begin{figure}[t]
	\centering
	\subfigure[]{\includegraphics[scale = 0.52]{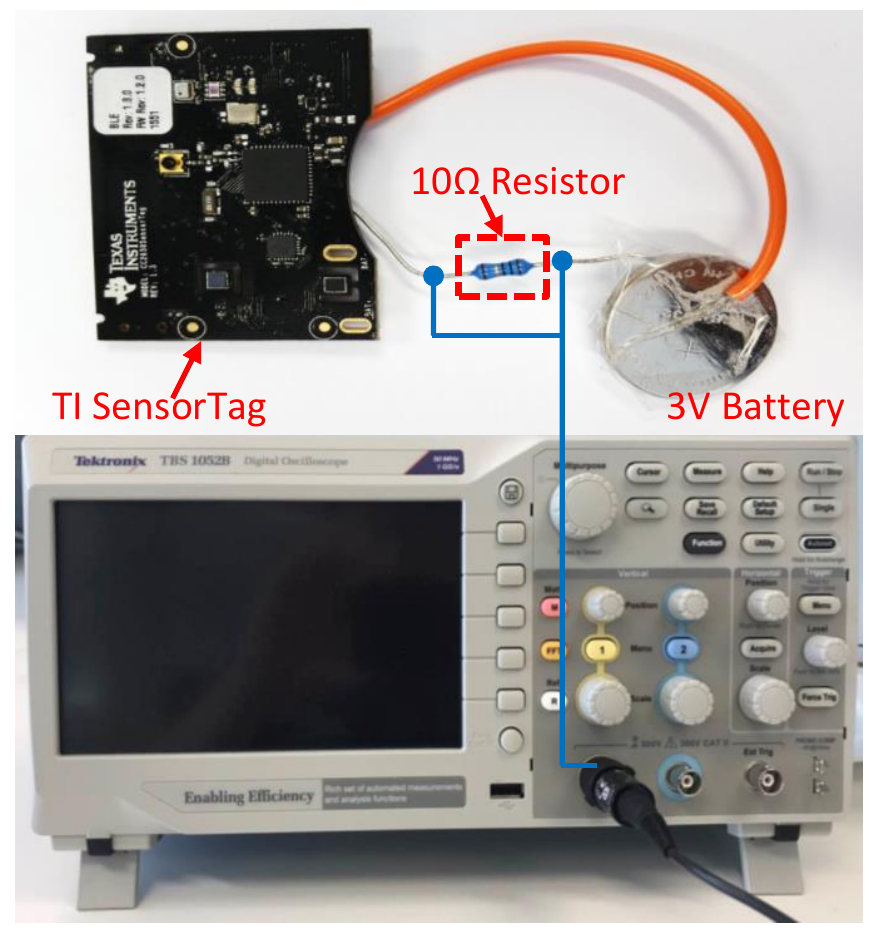}
		\label{fig:setup}}
	\subfigure[]{\includegraphics[scale = 0.52]{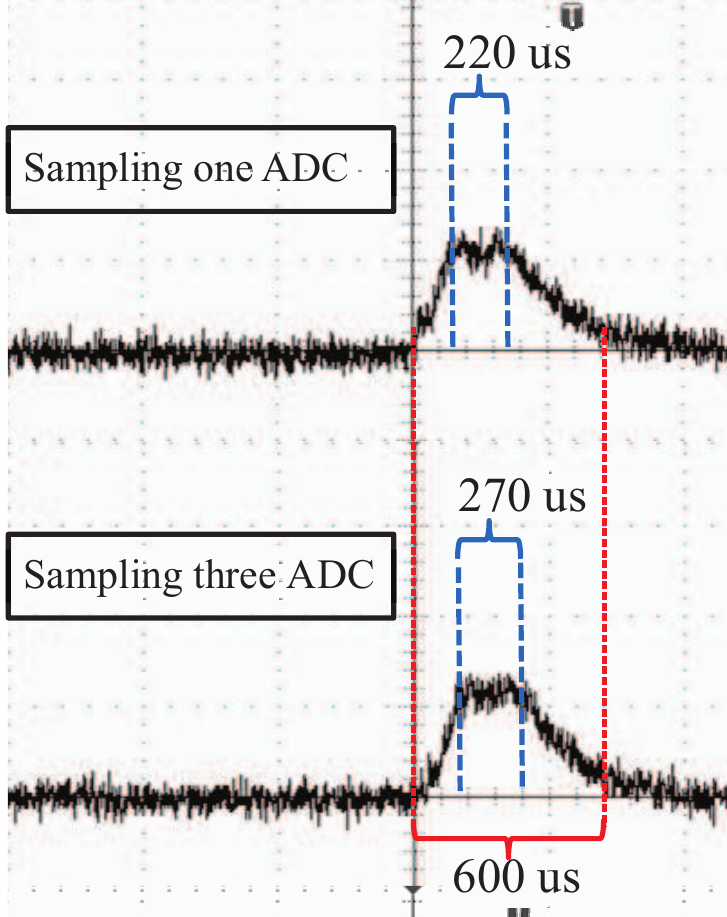}
		\label{fig:poweroverhead}}
	\caption{(a) Measurement setup, (b) Profiling of ADC sampling.}
	\vspace{-0.1in}
\end{figure}

Figure~\ref{fig:setup} shows the experiment setup. We select the Texas Instrument SensorTag as the target device, which is equipped with an ultra-low power ARM Cortex-M3 MCU and 12-bits ADC. The SensorTag is running with the Contiki 3.0 operating system and programs that periodically sample one and three ADC channels are loaded. By connecting a 10$\Omega$ resistor between the SensorTag and a 3V coin battery, we measure the voltage on the resistor using a Tektronix TBS-1052B digital oscilloscope thereby deriving the current draw by using the voltage divided by 10$\Omega$.

\vspace{-0.1in}
\subsubsection{Power consumption analysis}

Figure~\ref{fig:poweroverhead} illustrates the profiling of an ADC sampling event. At the beginning, the MCU is configured to deep sleep mode, which consumes around 6$\mu W$. Once the ADC sampling event is triggered, MCU wakes up and reads the value from ADC channels and then goes back to deep sleep mode. We can observe that the total time (refers to Total Event Time in Table\ref{tab:power}) of sampling one ADC channel and three ADC channels are almost equal at 600$\mu s$ (marked with red dashed line), while the major difference comes from the time duration (refers to MCU On Time) that the MCU is turned on (marked with blue dashed line). Note that the process of MCU turning on/off requires time, which refers to the climbing and dropping stage in the profiling. As shown in the figure, the MCU On Time is $220 \mu s$ and $270 \mu s$ when sampling one and three ADC channels respectively. We calculate the average power consumption (refers to ADC Power) during the total event time with the built-in function of the oscilloscope and the results are 482$\mu W$ and 511 $\mu W$ for one ADC and three ADC respectively.

Given a sampling rate of 40 Hz with duty-cycling, Table~\ref{tab:power} presents a detailed power consumption analysis comparing previous architecture\cite{xiang2013powering} (one ADC + amplifier) and the proposed SEHS architecture (three ADCs). First, we can see that sampling three ADC channels indeed incurs higher ADC Power but the increasement is not much. We calculate the consumed energy for each sampling event and find that only 0.02$\mu J$ additional energy is incurred, which suggests that the energy overhead of the proposed filter is subtle. Second, at 40 Hz sampling rate, we calculate the overall sensing power consumption when considering other components (e.g., amplifier) and MCU sleep time. Previous architecture consumes $29.43\mu W$\footnote{$P_{[6]} = (E_{ADC}+E_{Amp}+E_{Sleep})/t=(482\times600\times10^{-6}\times40+500\times600\times10^{-6}\times40+6\times(1-600\times40\times10^{-6}))/1 = 29.43 \mu W$.} while our SEHS architecture only consumes $18.12\mu W$\footnote{$P_{SEHS} = (E_{ADC}+E_{Sleep})/t=(511\times600\times10^{-6}\times40+6\times(1-600\times40\times10^{-6}))/1 = 18.12 \mu W$.}, achieve a power reduction of 38\%. These results suggest that SEHS not only achieves better system performance in terms of energy harvesting and sensing, but also decreases the sensing power consumption compared to the state-of-the-art.

\begin{table}
	\centering
	\caption{Power consumption analysis when sampling at 40 Hz.}
	\label{tab:power}
	\ra{1.3}
	\begin{tabular}{lcc}\toprule
		\textbf{Event}& \textbf{\cite{xiang2013powering} (One ADC)} & \textbf{SEHS (Three ADC)} \\
		\midrule
		
		\textbf{MCU On Time ($\mu s$)} &220  &270  \\ \midrule
		
		\textbf{Total Event Time ($\mu s$)} &600  &600  \\ \midrule
		
		\textbf{MCU Sleep Power ($\mu W$)}& 6 & 6 \\ 	\midrule	
		
		\textbf{ADC Power ($\mu W$)}& 482 & 511 \\ 	\midrule
		
		\textbf{Energy/Event ($\mu J$)}& 0.29 & 0.31 \\ 	\midrule
		
		\textbf{Amplifier Power ($\mu W$)}& 500 & 0 \\ 	\midrule

		\textbf{Overall Power ($\mu W$)}& 29.43 & 18.12 \\ 	
		\bottomrule
	\end{tabular}
	\vspace{-0.1in}
\end{table}
\vspace{-0.1in}
\section{Discussion}
\label{sec:discuss}
In this section, we discuss the limitations of current work and propose some potential solutions. First, the collected gait dataset came from volunteers walking only on soft ground (carpet) with normal speed. Prior investigations indicate that human gaits can be recognized with better accuracy for normal and fast walks compared to the case when the subject walks slowly~\cite{muaaz2012influence}, and hard walking ground enables higher accuracy than soft ground like grass~\cite{enokida2006predictive}.  Thus, a future work would be to collect SEHS data under more diverse walking grounds and speeds to asses the performance of SEHS more comprehensively.  Second, energy harvesting from foot strikes is promising due to its high energy density~\cite{8944276}, but the current work did not optimize energy harvesting efficiency. Techniques, such as maximum-power-point-tracking (MPPT)~\cite{subudhi2012comparative} could be used to maximize power extraction in shoe-based SEHS. Utilization of more advanced piezoelectric materials is another potential way to enhance energy generation~\cite{hassan2018kinetic}. Third, the number of subjects in our experiment is limited. It is expected that accurate gait detection will be more challenging when more subjects are involved. As a result, it may be necessary to explore more sophisticated deep learning models for SEHS-based gait recognition.

\vspace{-0.12in}
\section{Related work}
\label{section:related_work}
Our work relates to three main literature: context sensing using the output of PEH, recognition of human gait, and simultaneous energy harvesting and information sensing. We review these literature in this section.
\vspace{-0.1in}
\subsection{PEH based Context Sensing}
PEH possesses a well-known phenomenon called piezolectric effect: when subjected to an external force, a PEH device generates AC voltage of which amplitude is positively correlated to the intensity of force. With appropriate signal processing and pattern recognition algorithms, the PEH signal can be used as a proxy to sense and detect motion-related context. Based on that, researchers have demonstrated a wide range of context sensing applications using PEH. Han et al.~\cite{han2016self} built a shoe prototype with a PEH embedded. By analysing the generated AC waveforms, six different activities can be classified with over 90\% accuracy. Lan et al.~\cite{lan2019entrans} designed a wearable PEH prototype to measure the vibrations when taking different transportation facilities (e.g., car, bus, train, and ferry), which achieved around 85\% detection accuracy with typical machine learning. Other PEH based sensing applications include calorie expenditure estimation~\cite{lan2015estimating}, gait recognition~\cite{xuTMCGait,lin2020kehkey}, acoustic communication~\cite{lan2018hidden,lan2017veh}, and so on.

Compared to conventional motion sensor based approaches, PEH based method significantly reduces the sensing power consumption as PEH is a completely passive element~\cite{8944276}. However, unlike specialized sensors that are fine-tuned for accurate measurements of the physical phenomenon, PEH can be only regarded as a crude sensing device. As a result, the context detection performance is inferior to specialized sensors.

Unlike existing works that only consider the sensing capability of PEH, our work takes its original functionality (energy harvesting) into consideration as well. We have achieved simultaneous dual-use of a PEH and addressed the accompanying issues with signal processing. In addition, we introduce deep learning based classifiers to combat the low sensing capability of PEH.

\vspace{-0.1in}
\subsection{Human Gait Recognition}
Gait, the manner of human walking, has been recognized as a unique biometric feature of human. Recognizing human gait promises a number of applications such as people identification, device
	authentication, as well as health-related diagnostics like the Parkinson's disease~\cite{mirelman2019gait}. Researchers have employed various modalities to capture human gait. The vision-based method utilizes camera sensors to capture visual images of human body and extracts the temporal transition features of a specific body part (e.g., joints) within a gait cycle~\cite{zhao20063d}. The wireless-based method usually exploits RF signals (e.g., Wi-Fi~\cite{zhang2016wifi}, Radar~\cite{tahmoush2009radar}) or acoustic signal~\cite{xu2019acousticid} to extract a user's gait. The underlying rationale is that walking style of users leaves unique signature on the wireless signals. Unlike the above methods that require deployment of gait capturing facilities in the space, wearable-based gait recognition method entails users to wear a motion sensor (like accelerometer and gyroscope) so that the gait can be reflected on the pattern of motion signals. 

These methods have been proposed for a while and recent research focuses on the improvement of the recognition performance under more practical scenarios. Specifically, various deep neural networks have demonstrated their effectiveness in improving the recognition accuracy and robustness. For image data, Battistone et al.~\cite{battistone2019tglstm} proposed a graph based neural network (named Time based Graph Long Short-Term Memory, TGLSTM) to learn the long short-term dependencies of gait among frames. Chao et al.~\cite{zou2020deep} integrated the set perspective into a convolutional neural network to improve the robustness under diverse viewing angles or different clothes. For inertial sensor gait data, Zou et al.~\cite{chao2019gaitset} combined a convolutional neural network and a recurrent neural network to extract space and time domain features, which achieves 93\% accuracy among 118 subjects.

Although the feasibility of using PEH for human gait recognition has been demonstrated by Xu et al.~\cite{xuTMCGait}, our work differs in two aspects. First, instead of considering the sensing capability of PEH only, we proposed the simultaneous dual-use of PEH, identified the distortion effect of energy storage, and proposed a filter to resolve the effect. Second, Xu et al. exploits the hand-held PEH to capture vibrations during walking, while we utilized the foot-mounted PEH to capture the pressure of foot strikes. Our method is truly unobtrusive, insusceptible to irregular hand motions, and can harvest more energy. Compared to the conference version of this work~\cite{ma2018sehs}, we improved the gait recognition performance with deep learning.
\vspace{-0.1in}
\subsection{Simultaneous Energy Harvesting and Information Sensing}
With the sensing capability of energy harvesters being demonstrated, simultaneous energy harvesting and information sensing has attracted growing attention recently. For kinetic-based energy harvesting, instead of using additional ADC channels to minimize the impact of energy storage, Sandhu et al.~\cite{sandhu2020phd} proposed the use of a DC-DC boost converter to decouple the energy storage component from the transducer. However, the performance of this approach was not evaluated. For solar energy harvesting, Li et al.~\cite{li2018self} proposed a self-powered gesture recognition system using arrays of photodiodes. The photodiodes operate in photovoltaic mode to harvest energy from ambient light and their outputs are sampled to recognize different hand gestures.

In addition, the concept of SEHS is quite similar to simultaneous wireless information and power transfer (SWIPT) in wireless energy harvesting networks, where the RF signals are used for energy delivery as well as for information transmission.
	SWIPT has been extensively researched in the literature, including the typical structure~\cite{zhang2013mimo}, resource allocation strategy~\cite{lu2014wireless}, and energy-information trade-off~\cite{shi2014joint}, which have been surveyed in~\cite{lu2015wireless}. Most recently, researchers mainly focus on the application of SWIPT in advanced mobile networks, such as 5G networks where the use of millimetre wave brings new challenges~\cite{liang2019simultaneous,rajaram2019novel}, or unmanned aerial vehicle (UAV) networks where the mobility of UAVs affects the link quality dynamically~\cite{hong2019resource,wang2020joint}. 

\vspace{-0.1in}
\section{Conclusion}
\label{section:conclusion}

We have proposed SEHS, a novel architecture for simultaneous energy harvesting and gait recognition using the same piece of PEH hardware. To achieve high-accurate gait recognition, we proposed a filtering algorithm to minimize the sensing signal distortions caused by energy storage, and LSTM-based classifiers to mine useful information from noisy PEH data. We developed an insole-based prototype of SEHS and collected data from 20 subjects. Based on the experimental results, we have demonstrated that the SEHS prototype can harvest up to 127\% more energy and detect human gait with 12\% higher accuracy compared to the state-of-the-art. A power measurement confirms that SEHS achieves these performance improvements while actually consuming less power.
\vspace{-0.15in}



%

\bibliographystyle{IEEEtran} 
\bibliography{iotj}

\begin{IEEEbiography}[{\includegraphics[width=1in,height=1.25in,clip,keepaspectratio]{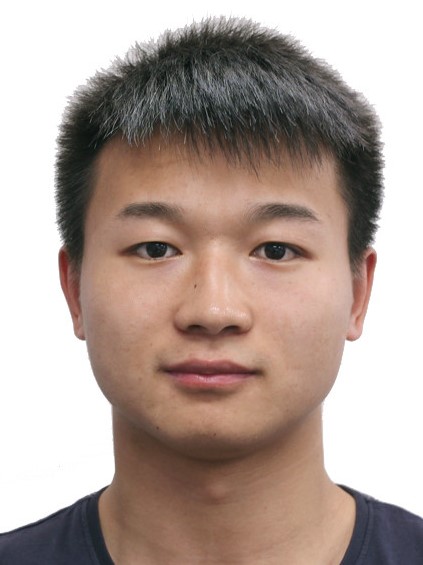}}]{Dong Ma} is a Ph.D. student in the School of Computer Science and Engineering, the University of New South Wales, Sydney, Australia. He received the B.S. degree in communication engineering from Central South University, China, in 2014. His research interests include Internet of Things, Pervasive Computing, Visible Light Sensing, and Energy Harvesting. He is a student member of the IEEE. 
\end{IEEEbiography}
\vspace{-0.2in}
\begin{IEEEbiography}[{\includegraphics[width=1in,height=1.25in,clip,keepaspectratio]{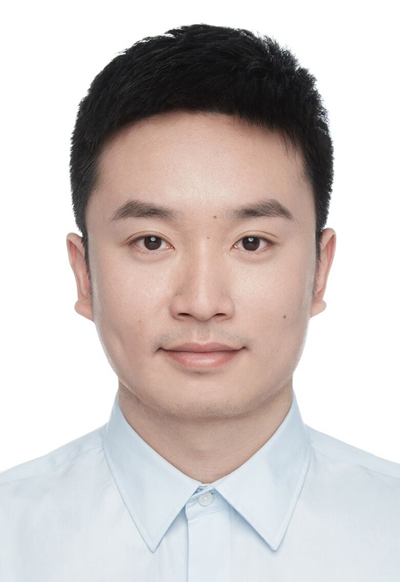}}]{Guohao Lan} is currently a Postdoctoral Research Associate with the Department of Electrical and Computer Engineering, Duke University, Durham, NC 27708, USA. Prior to this, he obtained his Ph.D. degree in Computer Science and Engineering from the University of New South Wales, Australia. He received the M.S. degree in Computer Science from Korea Advanced Institute of Science and Technology (KAIST), Korea, in 2015, and the B.E. degree in Software Engineering from Harbin Institute of Technology, China, in 2012. His research interests include wireless sensor networks, and mobile computing systems. He is a member of the IEEE.
\end{IEEEbiography}
\vspace{-0.25in}
\begin{IEEEbiography}[{\includegraphics[width=1in,height=1.25in,clip,keepaspectratio]{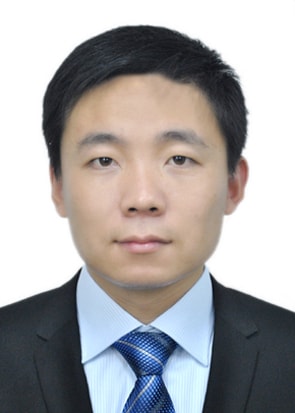}}]{Weitao Xu} is currently a Research Associate at the College of Computer Science and Software Engineering, Shenzhen University, China. He received his Ph.D. degree from the University of Queensland in 2017. He received the bachelor of engineering and the master of engineering degrees in 2010 and 2013, respectively, from the School of Information Science and Engineering, Shandong University, Shandong, China. He is a member of the IEEE.
\end{IEEEbiography}
\vspace{-0.2in}
\begin{IEEEbiography}[{\includegraphics[width=1in,height=1.25in,clip,keepaspectratio]{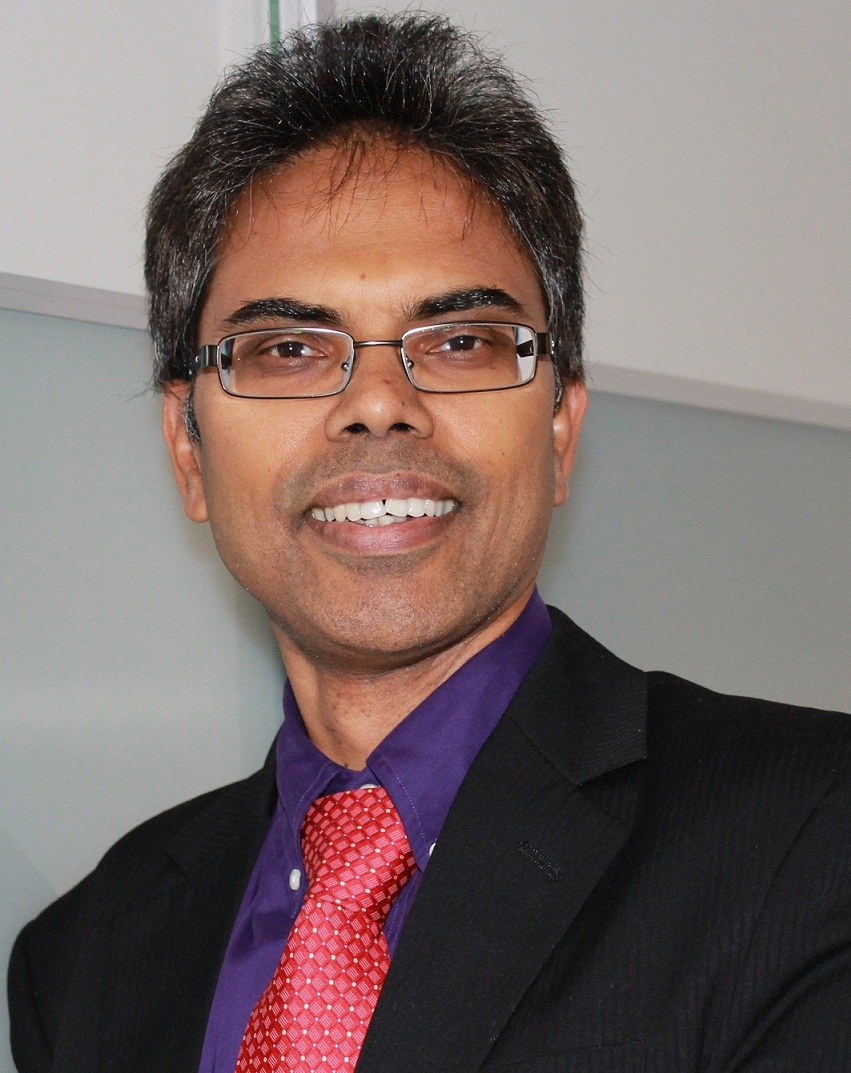}}]{Mahbub Hassan} is a Full Professor in the School of Computer Science and Engineering, the University of New South Wales, Sydney, Australia. He is a Senior Member of IEEE and served as a Distinguished Lecturer of IEEE (COMSOC) for 2013-2016. He is currently an Editor of IEEE Communications Surveys and Tutorial and has previously served as Guest Editor for IEEE Network, IEEE Communications Magazine, IEEE Transactions on Multimedia, and Area Editor for Computer Communications. Professor Hassan has earned a PhD from Monash University, Australia, and an MSc from University of Victoria, Canada, both in Computer Science. More information about Professor Hassan is available from http://www.cse.unsw.edu.au/~mahbub.
\end{IEEEbiography}
\vspace{-0.2in}
\begin{IEEEbiography}[{\includegraphics[width=1in,height=1.25in,clip,keepaspectratio]{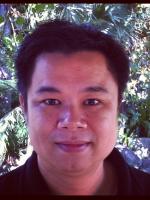}}]{Wen Hu} is an Associate Professor at School of Computer Science and Engineering, the University of New South Wales (UNSW). His research focuses on the novel applications, low-power communications, security and compressive sensing in sensor network systems and Internet of Things (IoT). Hu published regularly in the top rated sensor network and mobile computing venues such as ACM/IEEE IPSN, ACM SenSys, ACM transactions on Sensor Networks (TOSN), IEEE Transactions on Mobile Computing (TMC), Proceedings of the IEEE. Hu is a senior member of ACM and IEEE, an associate editor of ACM TOSN and the general chair of CPS-IoT Week 2020, as well as serves on the organising and program committees of networking conferences including ACM/IEEE IPSN, ACM SenSys, ACM MobiSys, ACM MobiCom, and  IEEE ICDCS.
\end{IEEEbiography}
\balance
\end{document}